\documentclass[aps,prd,twocolumn,showpacs,preprintnumbers,nofootinbib,amsmath,amssymb]{revtex4}

\usepackage{amsmath}
\usepackage{color}
\usepackage{bm}% bold math
\usepackage{dcolumn}% Align table columns on decimal point
\usepackage{graphicx}% Include figure files
\usepackage{multirow}

\newcommand{\beq}{\begin{eqnarray}}
\newcommand{\eeq}{\end{eqnarray}}
\newcommand{\ie}{{\it i.e. }}
\newcommand{\tr}{\ensuremath{\mathrm{Tr}}}

\newcommand{\ds}{\ensuremath{\displaystyle}}

\def\spose#1{\hbox to 0pt{#1\hss}}
\def\ltapprox{\mathrel{\spose{\lower 3pt\hbox{$\mathchar"218$}}
 \raise 2.0pt\hbox{$\mathchar"13C$}}}

\begin{document}

\title{Curvature of the {chiral} pseudo-critical line in QCD}

\author{Claudio Bonati}
\email{bonati@df.unipi.it}

\author{Massimo D'Elia}
\email{delia@df.unipi.it}

\author{Marco Mariti}
\email{mariti@df.unipi.it}

\author{Michele Mesiti}
\email{mesiti@pi.infn.it}

\author{Francesco Negro}
\email{fnegro@pi.infn.it}
\affiliation{
Dipartimento di Fisica dell'Universit\`a
di Pisa and INFN - Sezione di Pisa,\\ Largo Pontecorvo 3, I-56127 Pisa, Italy}

\author{Francesco Sanfilippo}
\email{f.sanfilippo@soton.ac.uk}
\affiliation{School of Physics and Astronomy, University of Southampton, SO17 1BJ Southampton, 
United Kindgdom}

\date{\today}% It is always \today, today,

\begin{abstract}
We determine the
curvature of the pseudo-critical line of strong interactions
by means of numerical simulations at imaginary chemical potentials.
We consider
$N_f=2+1$ stout improved staggered fermions with physical quark masses and
the tree level Symanzik gauge action,
and explore two different sets of lattice spacings, 
corresponding to temporal extensions
$N_t = 6$ and $N_t = 8$.  
Both the renormalized chiral condensate and the
renormalized chiral susceptibility are used to locate the transition. 
The determinations 
obtained from the two quantities are in good agreement, a preliminary 
continuum extrapolation yields $\kappa = 0.013(2)(1)$.
We also investigate the impact of a non-zero strange quark chemical potential and compare 
our results to previous determinations in the literature, 
discussing the possible sources 
of systematic errors affecting the various procedures.
\end{abstract}

\maketitle

\section{Introduction}
\label{intro}

The presence of a finite temperature transition to a new state of strongly
interacting matter, where chiral symmetry is restored and quarks and gluons are
deconfined, is a well established fact of Quantum Chromodynamics (QCD).
Present evidence from lattice QCD simulations indicates that a rapid change of
properties takes place, instead of a genuine phase transition, at a
pseudo-critical temperature $T_c$, which is around 155 MeV for chiral symmetry
restoration~\cite{aefks,afks,betal,tchot,tchot2}~\footnote{In the following,
however, for the sake of simplicity, we will refer to the 
rapid change of properties as the ``transition'', even if no critical 
behavior is associated with it.}.

How $T_c$ changes as a function of various external parameters is a question of
great importance, especially for considering QCD predictions in
phenomenological contexts where such parameters play a role. An example is the
baryon chemical potential, $\mu_B$, which enters the description of heavy ion
collisions and of {some} astrophysical objects.

A direct lattice determination of $T_c$ at $\mu_B \neq 0$ is presently hindered
by the well known sign problem, however various methods 
{have been proposed so far} to partially
circumvent it and {to} obtain 
{reliable results} in the regime of
small $\mu_B$.  A natural parametrization of $T_c (\mu_B)$ for small chemical
potentials, which exploits the symmetry under charge conjugation at $\mu_B = 0$
and assumes analyticity around {this point} is 
\begin{equation}\label{corcur}
\frac{T_c(\mu_B)}{T_c}=1-\kappa \left(\frac{\mu_B}{T_c}\right)^2\, +\, 
O(\mu_B^4)\, ,
\end{equation}
where the coefficient $\kappa$ defines the curvature of the pseudo-critical line
$T_c (\mu_B)$. {The curvature can be obtained in lattice QCD simulations in various ways:
by suitable combinations of expectation values computed at $\mu_B  = 0$ (Taylor
expansion method}~\cite{Allton:2002zi, Kaczmarek2011,
Endrodi2011,Borsanyi2012}); {by determining the pseudo-critical line for purely imaginary
values of $\mu_B$, for which numerical simulations are feasible, then fixing
the behavior for small and real $\mu_B$ by analytic
continuation}~\cite{fp1,dl1,adgl,wlc,ceasu2,ccdmp,ccdp,naganaka,nf2_ccdps,Laermann2013,ccp}; {by 
reweighting techniques, in which the oscillating complex behavior 
is shifted from the path integral measure to the physical observables}~\cite{FK1, FK2};
{by a reconstruction of the canonical partition function}~\cite{kf,afhl}.

A natural candidate for a comparison with QCD predictions is the chemical
freeze-out curve in the $T - \mu_B$ plane, determined by heavy ion collision
experiments, which is obtained so as to describe the particle
{multiplicities} according to a thermal-statistical
model~\cite{freezeout1, freezeout2, freezeout3, freezeout4, freezeout5,
freezeout6}.  Depending on the initial beam energy, different values of $T$ and
$\mu_B$ are obtained, which are thought to correspond to the conditions of last
chemical equilibrium of the thermal medium produced after the collision.  In
general, one can only assume that chemical freeze-out takes place after
re-hadronization, i.e.~that the freeze-out curve lies below the pseudo-critical
line in the $T - \mu_B$ plane.  However, it is also reasonable to guess that
chemical freeze-out is reached shortly after re-hadronization, so that the two
lines may lie close to each other.  Actually, a comparison with lattice
determinations of $T_c(\mu_B)$ does not show a good agreement.  In particular,
most determinations of $\kappa$ turn out be a factor 2-3 smaller than the
corresponding curvature of the freeze-out curve~\cite{Cleymans:2005xv}, even if
a recent reanalysis of experimental data, which takes into account inelastic
collisions after freeze-out, seems to bring to a significant reduction of such
a discrepancy~\cite{Becattini:2012xb}.

On the side of lattice QCD simulations, a complete control over all possible
systematics is also desirable.  That requires a proper continuum extrapolation
and a comparison among different methods adopted to partially overcome the sign
problem. Most determinations available at or around the
physical point (i.e.~by adopting quark masses tuned so as to yield a
physical hadron spectrum) have been obtained by the Taylor expansion
{method, see Ref.}~\cite{Kaczmarek2011} {(p4-improved action for staggered quarks) 
and Refs.}~\cite{Endrodi2011, Borsanyi2012} {(stout-smeared improved
action for staggered quarks)}. A recent determination~\cite{ccp},
obtained by the method of analytic continuation and adopting a HISQ/tree action
discretization of $N_f = 2+1$ QCD, has provided a value of the curvature which
is somewhat larger with respect to previous lattice determinations. 

One would like to fully understand the causes of this discrepancy.  Apart from
possible systematics lying behind either the Taylor expansion method or
analytic continuation, one should consider other standard sources of systematic
errors, among which the different possible definitions of $T_c$, the approach
to {the thermodynamical limit (finite size effects)}, to the continuum limit and to the physical point, and the different setup of
chemical potentials, $\mu_u$, $\mu_d$ and $\mu_s$, coupled respectively to the
up, down and strange quark numbers. For instance, the choice in Ref.~\cite{ccp}
has been to fix them to equal values ($\mu_u = \mu_d = \mu_s = \mu_B/3$), while
in previous determinations the choice has {typically} been $\mu_u = \mu_d = \mu_B/3$ and
$\mu_s = 0$, which is thought to better reproduce the thermal medium {in 
accordance with} 
the initial conditions of heavy ion collisions, which correspond to
strangeness neutrality.

All that claims for a {more} systematic investigation.  In the present
study we will provide a determination of the curvature by the method of
analytic continuation, adopting a {stout staggered fermions} discretization of $N_f = 2+1$ QCD with
physical {values of the quark masses, and} 
with the tree level improved
Symanzik gauge action. We will consider both the case $\mu_s = 0$ and
$\mu_s \neq 0$ and will monitor two different physical quantities{, namely} the
renormalized chiral condensate and its susceptibility{,} in order to locate the
pseudo-critical temperature for different values of the chemical potentials.
Two different sets of lattice spacings on lines of constant physics will be
considered (corresponding to $N_t = 6$ and $N_t = 8$, where $N_t$
is the temporal extent) in order to 
{estimate} ultraviolet (UV) cut-off effects. {For $N_t = 6$ we will consider also different spatial volumes in order to estimate finite size effects.}

Various purposes can be accomplished in this way:
\begin{enumerate}

\item we will compare determinations of the curvature obtained by different
methods but with the same lattice discretization~\cite{Endrodi2011};

\item we will compare determinations of the curvature obtained by the same
method (analytic continuation) but with different discretizations \cite{ccp}
(HISQ vs stout smeared staggered quarks);

\item we will obtain a first indication about the effects on the critical line
of the inclusion of a strange quark chemical potential.
\end{enumerate}

The paper is organized as follows. In Section~\ref{sec2} we provide some
details about the lattice discretization adopted in this study, as well as
about the observables chosen to locate $T_c$ and their renormalization.  In
Section~\ref{sec3} we describe the method of analytic continuation, with a
focus on the possible differences related to the inclusion (or non inclusion)
of a strange chemical potential $\mu_s$, which stem from differences in the
corresponding phase diagrams at imaginary chemical potentials. In
Section~\ref{sec4} we present our numerical results, which are 
compared to previous determinations in Section~\ref{sec5}, where we also
briefly review different methods to determine $\kappa$ adopted in the
literature. Finally, in Section~\ref{sec6}, we draw our conclusions.

\section{Numerical setup and observables}
\label{sec2}

For $N_f = 2+1$ QCD one can consider, in general, three independent chemical
potentials, $\mu_u, \mu_d$ and $\mu_s$, coupled respectively to $N_u, N_d$ and
$N_s$, i.e. the number of up, down and strange quarks. Different conventions
can be adopted, for instance it is usual to make reference to the conserved
charges $B$, $Q$ and $S$ (baryon number, electric charge and strangeness) and to
the chemical potentials coupled to them, $\mu_B$, $\mu_Q$ and $\mu_S$. The
conserved charges are related to the quark numbers by
\begin{eqnarray}
B &=& (N_u + N_d + N_s)/3 \nonumber \\
Q &=& (2 N_u - N_d - N_s)/3 \\
S &=& - N_s \nonumber
\label{defcharge}
\end{eqnarray}
from which the relations between the chemical potentials can be deduced
\begin{eqnarray}
\mu_u &=& \mu_B/3 + 2 \mu_Q/3 \nonumber \\
\mu_d &=& \mu_B/3 - \mu_Q/3 \\
\mu_s &=& \mu_B/3 - \mu_Q/3 -\mu_S \, .\nonumber
\label{defchem1}
\end{eqnarray}
In the following we make reference to the convention in terms of $\mu_u,\mu_d$
and $\mu_s$, and translate to the other convention when necessary (e.g., to
extract $\kappa$ given in Eq.~\eqref{corcur}).

We perform lattice simulations of $N_f=2+1$ QCD in the presence of purely
imaginary quark chemical potentials, $\mu_f=i\mu_{f,I},\ \mu_{f,I}\in\mathbb{R}$,
with $f=u,d,s$.  We consider the following euclidean partition function of the
discretized theory:
\begin{eqnarray}
\label{partfunc}
\mathcal{Z} &=& \int \!\mathcal{D}U \,e^{-\mathcal{S}_{Y\!M}} \!\!\!\!\prod_{f=u,\,d,\,s} \!\!\! 
\det{\left({M^{f}_{\textnormal{st}}[U,\mu_{f,I}]}\right)^{1/4}}
\hspace{-0.1cm}, \\
\label{tlsyact}
\mathcal{S}_{Y\!M}&=& - \frac{\beta}{3}\sum_{i, \mu \neq \nu} \left( \frac{5}{6}
W^{1\!\times \! 1}_{i;\,\mu\nu} -
\frac{1}{12} W^{1\!\times \! 2}_{i;\,\mu\nu} \right), \\
\label{fermmatrix}
(M^f_{\textnormal{st}})_{i,\,j}&=&am_f \delta_{i,\,j}+\!\!\sum_{\nu=1}^{4}\frac{\eta_{i;\,\nu}}{2}\nonumber
\left[e^{i a \mu_{f,I}\delta_{\nu,4}}U^{(2)}_{i;\,\nu}\delta_{i,j-\hat{\nu}} \right. \nonumber\\
&-&\left. e^{-i a \mu_{f,I}\delta_{\nu,4}}U^{(2)\dagger}_{i-\hat\nu;\,\nu}\delta_{i,j+\hat\nu}  \right] \, .
\end{eqnarray}
The functional integration is performed over the gauge links, $U\in SU(3)$ and
$\mathcal{S}_{Y\!M}$ is the tree level improved Symanzik action introduced in
Refs.~\cite{weisz,curci}, where $W^{n\!\times \! m}_{i;\,\mu\nu}$ is the trace of the
$n\times m$ loop constructed from the gauge links along the directions $\mu,
\nu$ departing from the $i$ site.  The staggered Dirac operator
$(M^f_{\textnormal{st}})_{i,\,j}$ defined in Eq.~(\ref{fermmatrix}) is built up
in terms of the two times stout-smeared links $U^{(2)}_{i;\,\nu}$, which are
constructed following the procedure described in Ref.~\cite{morning} and using
the isotropic smearing parameters {$\rho_{\mu\nu}=0.15$ for $\mu\neq \nu$}.
The
stout smearing improvement is used in order to reduce the effects of finite
lattice spacing and taste symmetry violations (for a comparison between
different improved staggered discretizations and their effectiveness in
reducing taste violations, see Ref.~\cite{bazavov}). For each flavor we
introduce the imaginary quark chemical potential in the Dirac operator by
multiplying all the temporal links in the forward (backward) direction by $e^{+
i a \mu_{f,I}}$ ($e^{- i a \mu_{f,I}}$): that can be viewed as a rotation by an
angle $\theta_f \equiv a N_t \mu_{f,I} = \mu_{f,I}/T$ of the temporal boundary
conditions for the quark flavor $f$. For each lattice we have studied 3-4
different values of the imaginary chemical potentials (see
Table~\ref{tab:tc_all}).
As usual for the staggered fermions simulations, the residual fourth degeneracy
of the lattice Dirac operator is removed by using the rooting procedure:
{possible systematics related to this approach have been discussed
in the literature (see, e.g., Ref.~\cite{rooting}) and are shared
by the other lattice studies with which we are going to 
compare}~\cite{Kaczmarek2011, Endrodi2011, Borsanyi2012, ccp}.

We perform simulations at finite temperature around the transition temperature,
using lattices with two different temporal extensions, $N_t=6,8$. At fixed
$N_t$, the temperature $T=1/(aN_t)$ of the system is changed by varying
the value of the bare coupling constant $\beta$.  The bare quark masses $m_s$
and $m_{l}$ are accordingly rescaled with $\beta$, in order to move on a line of
constant physics, with $m_{\pi}\simeq 135\,\mathrm{MeV}$ and
$m_s/m_{l}=28.15$; this line is determined by a spline
interpolation of the values reported in Refs.~\cite{tcwup1,befjkkrs},
{the complete set of adopted parameters is reported in 
Appendix~\ref{app:t0}}.
{At fixed $T$, the two different values of $N_t$ correspond 
to different lattice spacings and permit us to estimate 
UV cut-off effects.}

The chiral condensate of the flavor $f$ is defined as
\begin{equation}
\langle\bar{\psi}\psi\rangle_f=\frac{T}{V}\frac{\partial \log Z}{\partial m_f}\ ,
\end{equation} 
where $V$ is the spatial volume. Since in our simulations the two light quarks 
are degenerate with mass $m_l\equiv m_u=m_d$, it is convenient to introduce the
light quark condensate:
\begin{equation}
\langle\bar\psi\psi\rangle_l=\frac{T}{V}\frac{\partial \log Z}{\partial m_l}=
\langle\bar{u}u\rangle+\langle\bar{d}d\rangle\ ,
\end{equation} 
which will be renormalized by adopting the prescription introduced
in Ref.~\cite{Cheng:2007jq}:
\begin{equation} \label{rencond}
\langle\bar{\psi}\psi\rangle^r_{l}(T)\equiv\frac{\left[
\langle \bar{\psi}\psi\rangle_l -\frac{\ds 2m_{l}}{\ds m_s}\langle \bar{s}s\rangle\right](T)}{
\left[\langle \bar{\psi}\psi\rangle_l-\frac{\ds 2m_{l}}{\ds m_s}\langle \bar{s}s\rangle\right](T=0)}\ ,
\end{equation} 
where $m_s$ is the bare strange quark mass.

The light quarks chiral susceptibility is given by ($M_l$ is the 
{Dirac operator} corresponding to a single light flavor)
\begin{eqnarray}
\label{susc}
\chi_{\bar\psi\psi}&=& \frac{\partial\langle\bar\psi\psi\rangle_l}{\partial m_l}
=\chi_{\bar\psi\psi}^{disc}+\chi_{\bar\psi\psi}^{conn}\\
\label{sconn}
\chi_{\bar\psi\psi}^{disc}&\equiv&\frac{T}{V}\left(\frac{N_l}{4}\right)^2
\left[\langle (\tr M_l^{-1})^2\rangle-\langle \tr M_l^{-1}\rangle^2 \right] \\
\label{conn}
\chi_{\bar\psi\psi}^{conn}&\equiv&
-\frac{T}{V}\frac{N_l}{4}\langle \tr M_l^{-2}\rangle\,.
\end{eqnarray}
In this expression $N_l$ is the number of degenerate light quarks, 
{that} in our case is fixed to $N_l=2$. Traces
are computed by noisy estimators, with $8$ random vectors for each flavor. 
The renormalization of the chiral susceptibility is performed by first
subtracting the $T=0$ contribution (thus removing the additive renormalization)
and then multiplying the result by the square of the bare light quark mass to
fix the multiplicative UV divergence~\cite{tcwup1}:
\begin{equation} \label{rensusc}
\chi_{\bar{\psi}\psi}^r=m_{l}^2\left[ \chi_{\bar{\psi}\psi}(T)-\chi_{\bar{\psi}\psi}(T=0)\right]\,.
\end{equation} 
All the $T=0$ quantities have been measured on a symmetric 
$N_t=N_s=32$ lattice.

The renormalization prescriptions Eqs.~\eqref{rencond}-\eqref{rensusc} are not
the only available choices: other approaches exist in the literature (see e.g.
Refs.~\cite{Kaczmarek2011, Endrodi2011}) and in the following sections we will
also investigate the dependence of the results on the different renormalization 
prescriptions adopted.

\section{Analytic continuation with and without a strange quark 
chemical potential}
\label{sec3}

Both the method of analytic continuation from imaginary chemical potentials and
the Taylor expansion approach are based on the assumption that the free energy
{is analytic, 
at least in a limited region of small chemical potentials.}

As it happens for other thermodynamical quantities, it is possible to make an
ansatz for the dependence of the pseudo-critical temperature
$T_c(\mu_u,\mu_d,\mu_s)$ which is valid for small chemical potentials.  The
symmetries of the theory constrain the possible form of this dependence. {First of all,}
 charge conjugation symmetry imposes that $T_c$, as well as the free
energy itself, be invariant under a simultaneous sign
change of all chemical potentials, thus
a Taylor expansion of $T_c$
must include only {monomials of overall even degree in the chemical
potentials.}

Moreover, in the case of two degenerate flavors, isospin symmetry imposes further
constraints. 
{By rewriting the coupling to the chemical potentials of the continuum lagrangian in the more compact form}
$\bar{\psi}_{f}\gamma_0M_{ff'}\psi_{f'}$, where $f, f'$
are flavor indices and $M$ is a $2\times 2$ hermitian matrix, {which is usually
diagonal}, it can be shown that the theory is invariant under {isospin}
{rotations} $\psi\to
R\psi$, combined with $M\to RMR^{\dag}$, where $R$ denotes a generic $SU(2)$
matrix.  The dependence of $T_c$ on the chemical potentials must satisfy such
invariance, that means that it can be function only of invariant quantities of
the matrix $M$. At the leading quadratic order two independent such
quantities exist, which can be chosen to be $\mathrm{det}(M)$ and
$\mathrm{Tr}(M^{\dag}M)$ (or alternatively $[\mathrm{Tr}(M)]^2$).  To leading
order we thus have {(see also Ref.~\cite{ds1})}:
\begin{equation}\label{quadratic_nf2}
\begin{aligned}
T_c(M)&=T_c(0)-\alpha \mathrm{Tr}(M^{\dag}M)-\gamma\, \mathrm{det}(M)=\\
&=T_c(0)-\alpha(\mu_u^2+\mu_d^2)-\gamma\, \mu_u\mu_d \, .
\end{aligned}
\end{equation}
It is interesting that at this level the requirement of isospin invariance is
in fact equivalent to the requirement of  symmetry under $u \leftrightarrow d$
exchange.  This is particularly relevant since in the lattice discretization of
the partition function, Eq.~\eqref{partfunc}, each quark is treated by means of
a separate quartic root of a fermion determinant, so that only the symmetry
under $u \leftrightarrow d$ exchange is strictly true at finite lattice
spacing.  It has been verified in Ref.~\cite{nf2_ccdps} that, in the
discretized $N_f = 2$ theory, the mixing term $\gamma$ is small but non-zero,
corresponding to a measurable difference between the curvature in terms of the
baryon or the isospin chemical potential.  In our $N_f=2+1$ setup, the
generalization of Eq.~\eqref{quadratic_nf2} is simply
\begin{eqnarray}
T_c(\mu_u,\mu_d,\mu_s) &=& T_c(0) - A (\mu_u^2 + \mu_d^2) - B \mu_s^2 \nonumber \\
-\ C \mu_u \mu_d  &-& D \mu_s (\mu_u + \mu_d)
 + O(\mu^4) \, .
\label{quadratic_exp}
\end{eqnarray}

In this study, we are interested only in two particular setups of chemical
potentials.  In the first case we set $\mu_u = \mu_d \equiv \mu_{l}$ and $\mu_s
= 0$.  That coincides with the setup adopted in most studies (like e.g. in
Refs.~\cite{Kaczmarek2011, Endrodi2011}), which is thought to be close to the
situation created in heavy ion collisions, where the initial conditions
correspond to strangeness neutrality. In this case the expected parametrization
is 
\begin{equation}
T_c(\mu_{l}) = T_c(0) - A' \mu_{l}^2 + O(\mu_{l}^4) 
\label{quadratic2}
\end{equation}
where $A' = A + C$. 

In the second case, we set $\mu_s = \mu_{l}$: that permits to estimate the
effects of the inclusion of $\mu_s$ and to compare with some previous
studies~\cite{ccp,Laermann2013}.  In this case the parametrization is 
\begin{equation}
T_c(\mu_{l}) = T_c(0) - (A' + B') \mu_{l}^2 + O(\mu_{l}^4) 
\label{quadratic3}
\end{equation}
where $B' = B + 2D$. An independent study with $\mu_u = \mu_d = 0$ and $\mu_s
\neq 0$ would provide direct information on $B$ and verify if the mixing term
$D$ is negligible or not: this is left for future investigations.

When the chemical potentials are purely imaginary, and if analytic
continuation holds true, the following dependence is expected for $T_c$ as a
function of the quantity $\theta_{l} = {\rm Im} (\mu_{l})/T$
{introduced in Section~\ref{sec2}}:
\begin{equation}
\frac{T_c(\theta_{l})}{T_c(0)} = 1 + R\, \theta_{l}^2 + O(\theta_{l}^4) 
\label{quadratic4}
\end{equation}
where $R = A'\, T_c(0)$ or $R = (A' + B')\, T_c(0)$, depending on the different
setup adopted.

\begin{figure}[t!]
\includegraphics[width=0.92\columnwidth, clip]{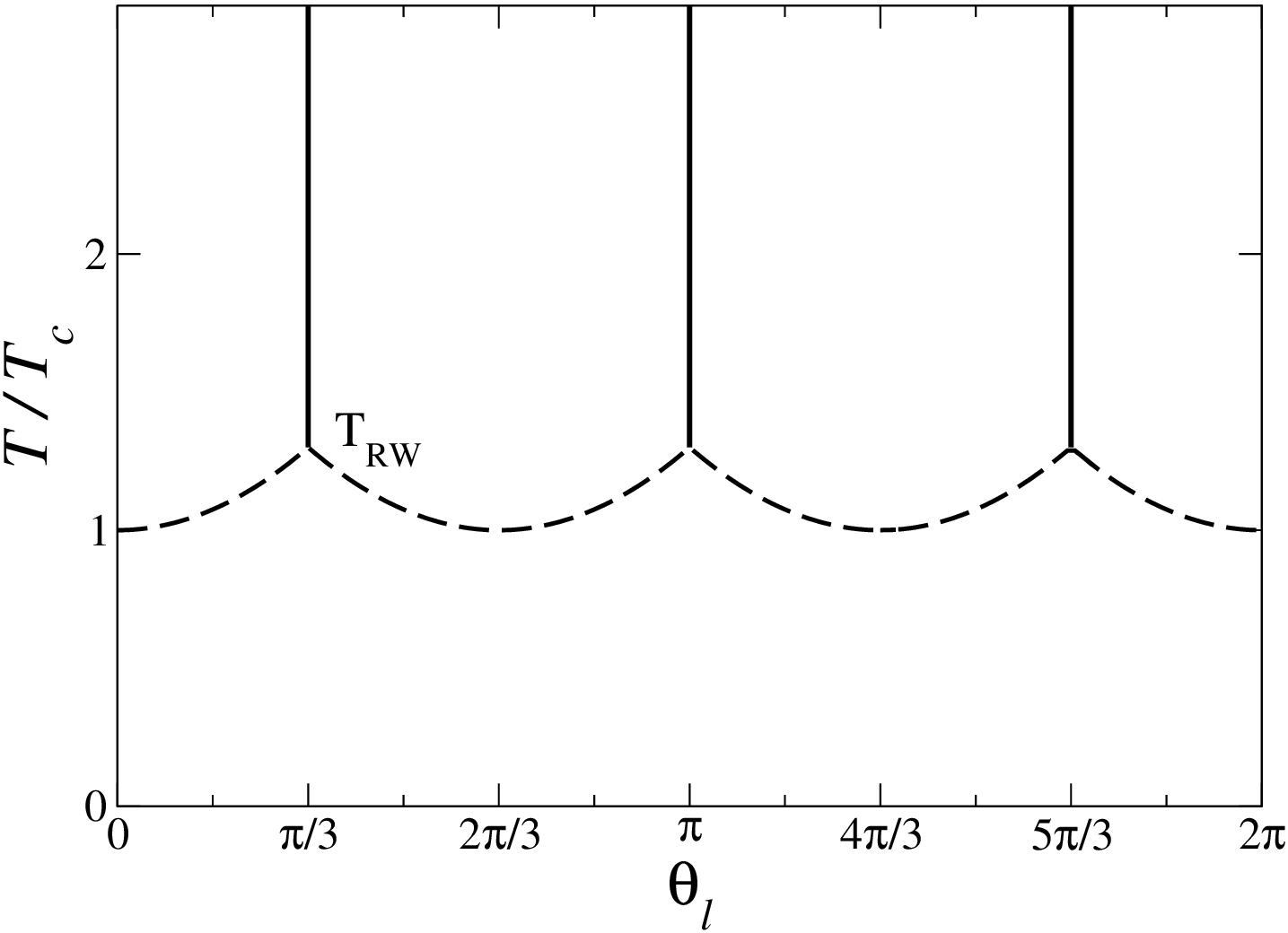}
\caption{Sketched phase diagram in the $T - \theta_l$ plane for $\mu_s =
\mu_{l}$. Solid lines indicate the RW lines, while the dashed lines corresponds
to the analytic continuation of the pseudo-critical line.}
\label{immu1}
\end{figure}

\begin{figure}[t]
\includegraphics[width=0.92\columnwidth, clip]{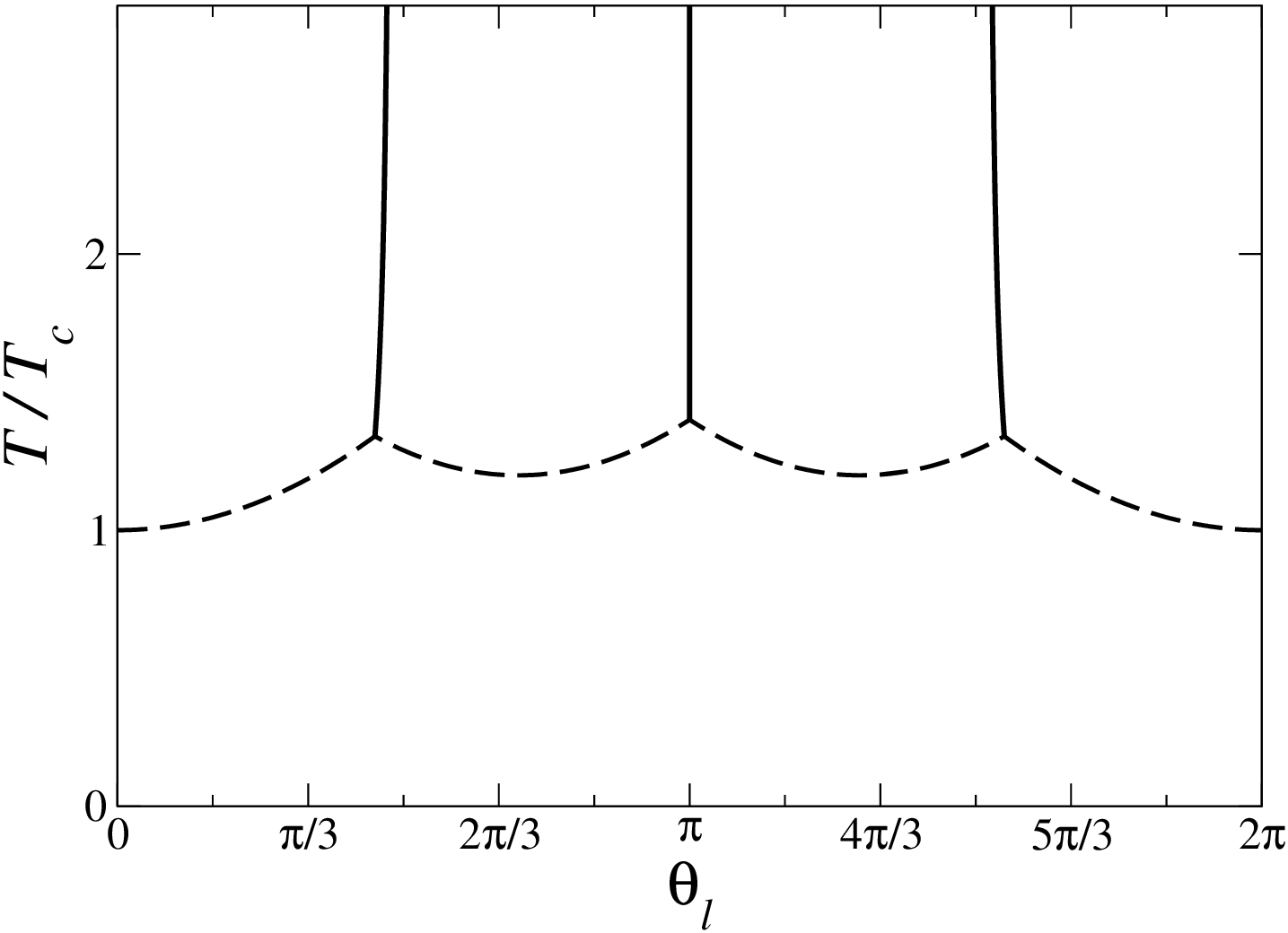}
\caption{The same as for Fig.~\ref{immu1}, but for $\mu_s = 0$. 
{In this case the exact location of the RW-like lines, apart from the 
one at $\theta_l = \pi$, is temperature dependent and known analytically
only in the high $T$ limit.}}
\label{immu2}
\end{figure}

Apart from the possible different values of the curvatures, the fact that $\mu_s = 0$ or
$\mu_s = \mu_{l}$ is of course relevant also to non-linear terms in
$\theta_{l}^2$, 
{as we are going to discuss in the rest of this section}.  
In this respect, a substantial difference in the phase diagram
in the $T - \theta_{l}$ plane may play a significant role. It is well
known~\cite{RW} that when all imaginary chemical potentials are equal, i.e.
when the temporal boundary conditions of all quark fields are rotated by the
same angle $\theta_{l}$, a translation of $\theta_{l}$ by a multiple of $2 \pi
/3$ can be cancelled by a center transformation of gauge fields, so that the
partition function is periodic in $\theta_{l}$ with a period $2 \pi /3$.  Such
a periodicity is smoothly realized at low $T$ \cite{fp1, dl1, ddl}, while in
the high $T$ regime it is enforced by first order transitions \cite{RW}, known
as Roberge-Weiss (RW) transitions, which are connected with center symmetry and
with the dynamics of the Polyakov loop, as explained in more details in the following.

In absence of dynamical fermions, the theory is invariant under center
transformations, i.e. gauge transformations which are periodic in time up to a
center phase $\exp(i 2 \pi k /3)$, where $k$ is an integer. However, the trace
of the Wilson line in the time direction (Polyakov loop) is not invariant
{under such transformations}.  A non-zero expectation value of the Polyakov
loop signals the spontaneous breaking of center symmetry at high $T$, where the
free energy develops three degenerate minima, corresponding to a Polyakov loop
oriented along the three different roots of unity $\exp(2 \pi i k/3)$, $k =
0,1,2$.  The fermion determinant breaks center symmetry explicitly: for zero
imaginary chemical potential its effect is like adding an external magnetic
field in a three-dimensional Potts model, aligning the Polyakov loop along the
positive real axis in the complex plane, i.e. along $k = 0$ (see e.g.
\cite{potts}).  An imaginary chemical potential coupled to quark flavor $f$, by
rotating the temporal boundary conditions in the fermion determinant, rotates
the coupling to the Polyakov loop by an angle $-\theta_f = -{\rm Im}
(\mu_f)/T$.

In the high $T$ phase, if all imaginary chemical potentials are set equal,
the Polyakov loop undergoes an abrupt change of orientation, corresponding to
the RW transitions, as $\theta_{l}$ crosses $\pi/3$ or odd multiples of it.
For such values, the effective magnetic-like coupling to the Polyakov field
points exactly in-between two cubic roots of unity, thus leaving an exact
residual $Z_2$ center symmetry, which is spontaneously broken at high $T$. The
onset of this spontaneous breaking, taking place at the endpoint of the
RW line, or RW endpoint, has been the subject of various
studies {(for a collection of lattice investigations, see Refs.}~\cite{DES, BCDES, PP_wilson, OPRW, nakamura, alexandru, wumeng, nknn}).  

The presence of the RW transitions places a limitation on the region of
imaginary chemical potentials available to analytic continuation: for high $T$,
only chemical potentials such that $\theta_{l} < \pi/3$ can be used to
investigate the dependence of the free energy for small values of $\mu_{l}$,
since for $\theta_{l} > \pi/3$ one is exploring a different analyticity 
sheet of
the free energy, corresponding to a different center sector, even if with
identical and periodically repeated {physical} properties.  The pseudo-critical line
itself, in particular, develops a non-analyticity at $\theta_{l} = \pi/3$:
numerical evidence is that it touches the RW endpoint, where it forms a cusp,
and then repeates periodically; such a situation is depicted schematically in
Fig.~\ref{immu1}. 

\begin{figure}[t!]
\includegraphics[width=0.92\columnwidth, clip]{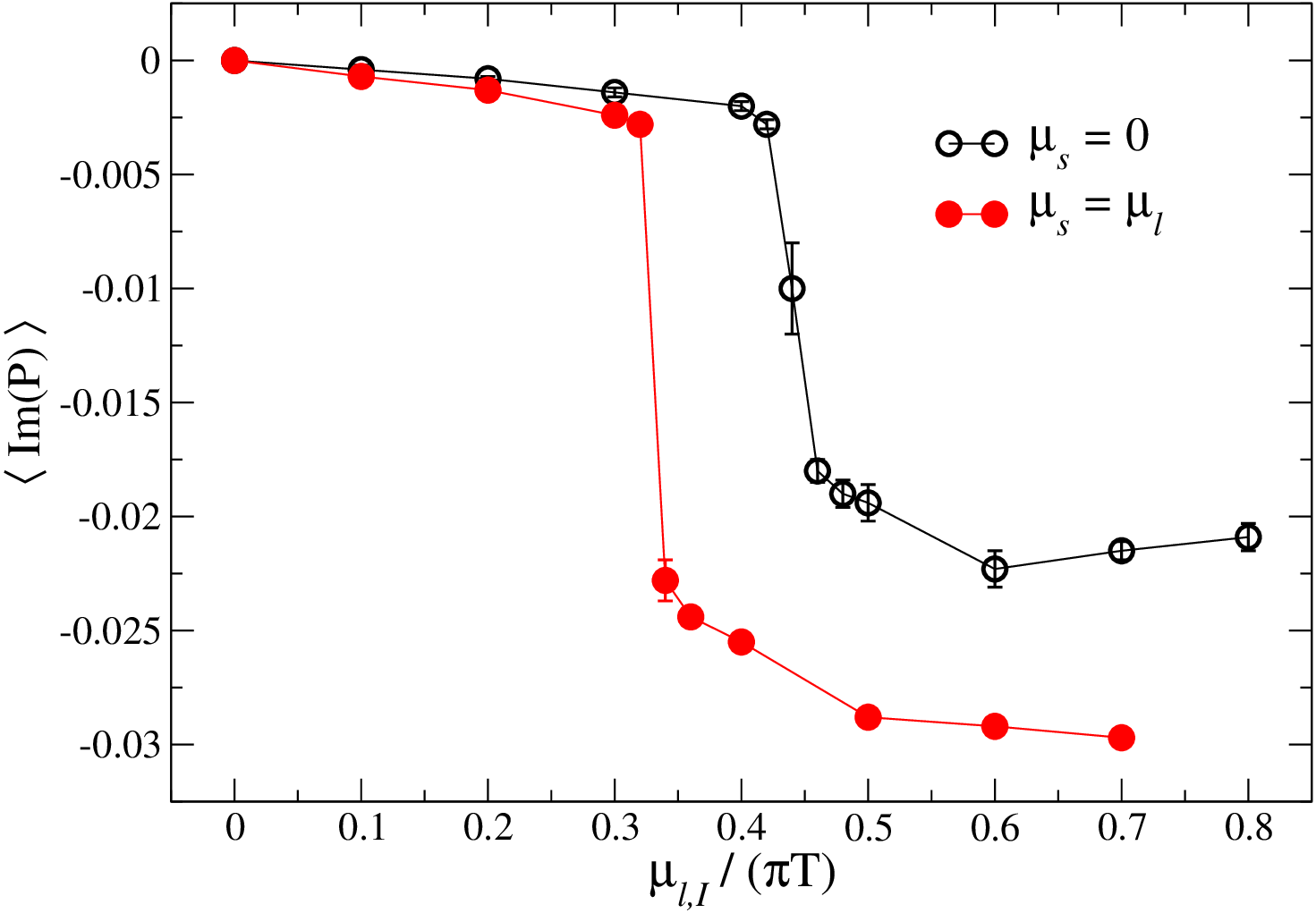}
\caption{Imaginary part of the Polyakov loop as a function of $\theta_l$ at
fixed $T \approx 208\,\mathrm{MeV}$ for $\mu_s = \mu_l$ and for $\mu_s  = 0$.}
\label{impol}
\end{figure}

When one adopts the setup in which $\mu_d = \mu_u \equiv \mu_{l}\neq 0$ and
$\mu_s = 0$, the phase diagram in the $T - \theta_{l}$ plane looks different.
The strange quark determinant is independent of $\theta_{l}$ and that breaks
the $\pi/3$ periodicity in $\theta_{l}$.  In particular, as $\theta_{l}$ is
increased, the effective coupling of the up and down quark determinants to the
Polyakov loop will rotate by an angle $-\theta_{l}$, while that of the strange
quark will stay oriented along the positive real axis. As a consequence, the
critical value of $\theta_{l}$ at which, in the high $T$ regime, the Polyakov
loop jumps from one sector to the other, will be higher than $\pi/3$.  Given
the residual $2 \pi$ periodicity in $\theta_{l}$ and the symmetry under
inversion of $\theta_{l}$, the expected phase diagram is depicted schematically
in Fig.~\ref{immu2}: we still have RW-like transition lines at high $T$, which
however take place for different values of $\theta_{l}$ (apart from the one at
$\theta_{l} = \pi$) and separate sectors of the theory which are not equivalent
to each other. 

We have verified this expectation explicitly by monitoring the Polyakov loop as
a function of $\theta_{l}$ in the two different setups: results are reported in
Fig.~\ref{impol}, where we plot the imaginary part of the Polyakov loop (which
jumps when the boundary between two different center sectors is crossed) as a
function of $\theta_{l}$.  While for $\mu_s = \mu_{l}$ the jump takes place at
$\theta_{l} = \pi/3$, as expected, when $\mu_s = 0$ the jump moves forward and
takes place approximately at $\theta_{l} \simeq 0.45\, \pi$.  {A
perturbative computation in the regime of asymptotically high temperatures,
performed making use of the one loop effective potential for the Polyakov loop
in the presence of massless quarks, gives $\theta_c\approx 0.482933\, \pi$, see
Appendix~\ref{appendix:effective_poly}.}

One important consequence is that, for high $T$, the region available for
analytic continuation is larger for $\mu_s = 0$ than for $\mu_s = \mu_{l}$:
that means that a better control on systematic effects can be attained.  Since
analytic continuation is actually performed in terms of $\theta_{l}^2$, going
from $\pi/3$ to approximately $0.45\, \pi$ means that the available region is
almost doubled, i.e. the increase is substantial. Moreover, one may expect that
{for $\mu_s=0$} the possible effects of the critical behavior around the
RW endpoint on the region of small chemical potentials should be milder, since
the endpoint is moved further inside the $T - \theta_{l}$ plane: such effects
include the possible non-linear contributions in $\theta_{l}^2$ to the
pseudo-critical line $T_c(\theta_{l})$. This point will be discussed further in
Section~\ref{sec4}.

\section{Numerical Results}
\label{sec4}

We performed simulations for different values of the chemical potentials and
$\mathcal{O}(10)$ temperatures around $T_{c}(\mu)$, {on four different lattice
sizes: $16^3\times6$, $24^3\times6$, $32^3\times6$ and $32^3\times8$}.  We mainly considered
the $\mu_s=0$ setup, for the $32^3\times8$ lattice we also studied the case
$\mu_s=\mu_{l}$.  The Rational Hybrid Monte-Carlo algorithm~\cite{rhmc1, rhmc2,
rhmc3} has been used for sampling gauge configurations according to
Eq.~(\ref{partfunc}), each single run consisting of 2-5 K trajectories of unit
length in molecular dynamics time, with higher statistics around the
transition.  

To perform the renormalization described in Sec.~\ref{sec2}, one needs to
compute observables also at $T=0$ and at the same values of the bare
parameters, i.e. at the same UV cutoff.  At $T = 0$ observables  depend
smoothly on $\beta$; moreover no dependence at all is expected on the imaginary
chemical potentials, since they can be viewed as a modification in the temporal
boundary conditions which, at $T = 0$ (hence for ideally infinite temporal
extension), are completely irrelevant.  For those reasons, we performed a set
of simulations on a $32^4$ lattice, at zero chemical potentials and for a few
values of $\beta$ on the line of constant physics. Then we estimated the
observables at intermediate values of $\beta$ by a suitable interpolation and
adopted them to renormalize data at $T \neq 0$ and generic values of the
chemical potentials. {More details on the $T = 0$ measurements are reported in
Appendix~\ref{app:t0}.}

\begin{figure}[h!]
\includegraphics[width=0.92\columnwidth, clip]{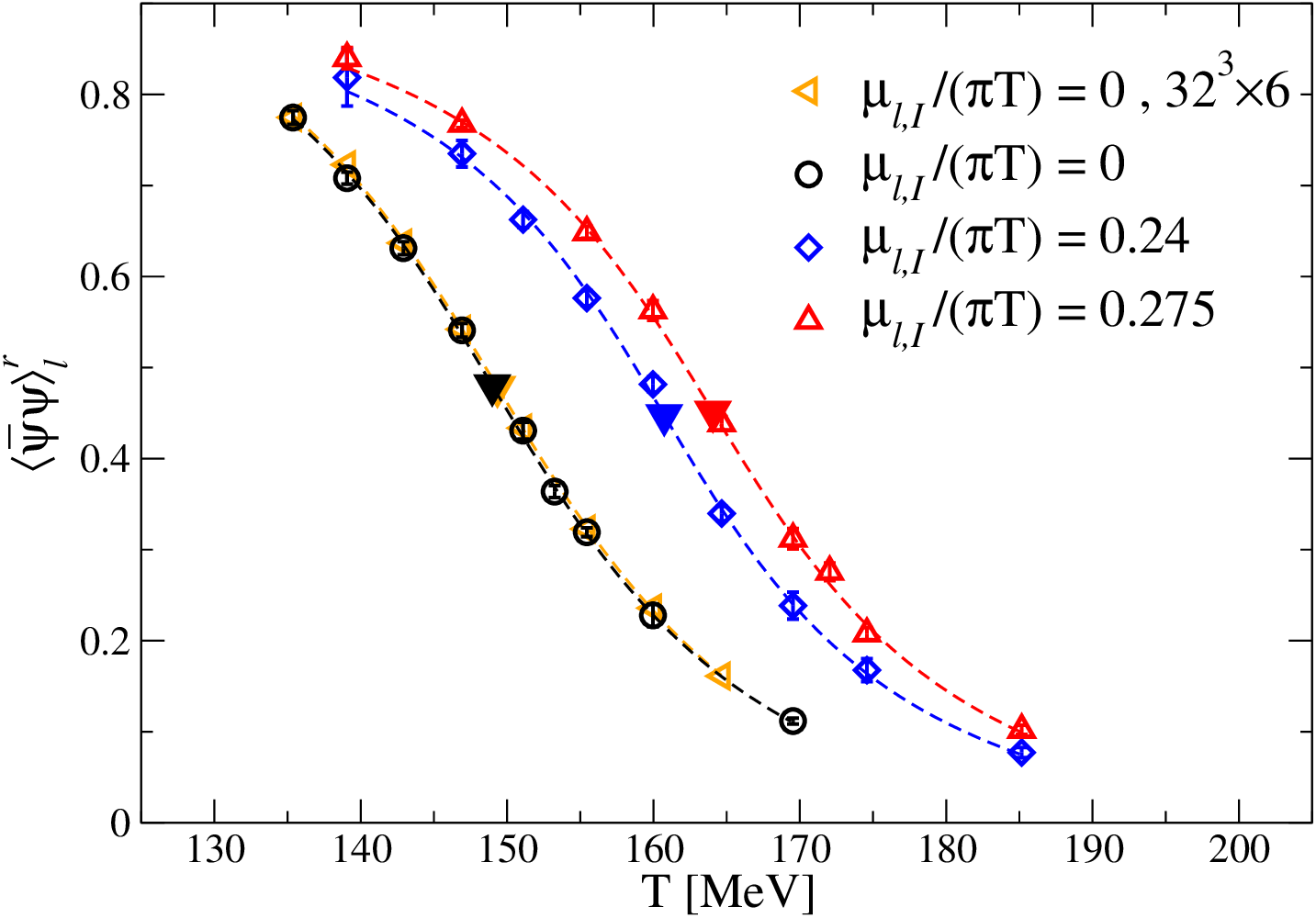}
\includegraphics[width=0.92\columnwidth, clip]{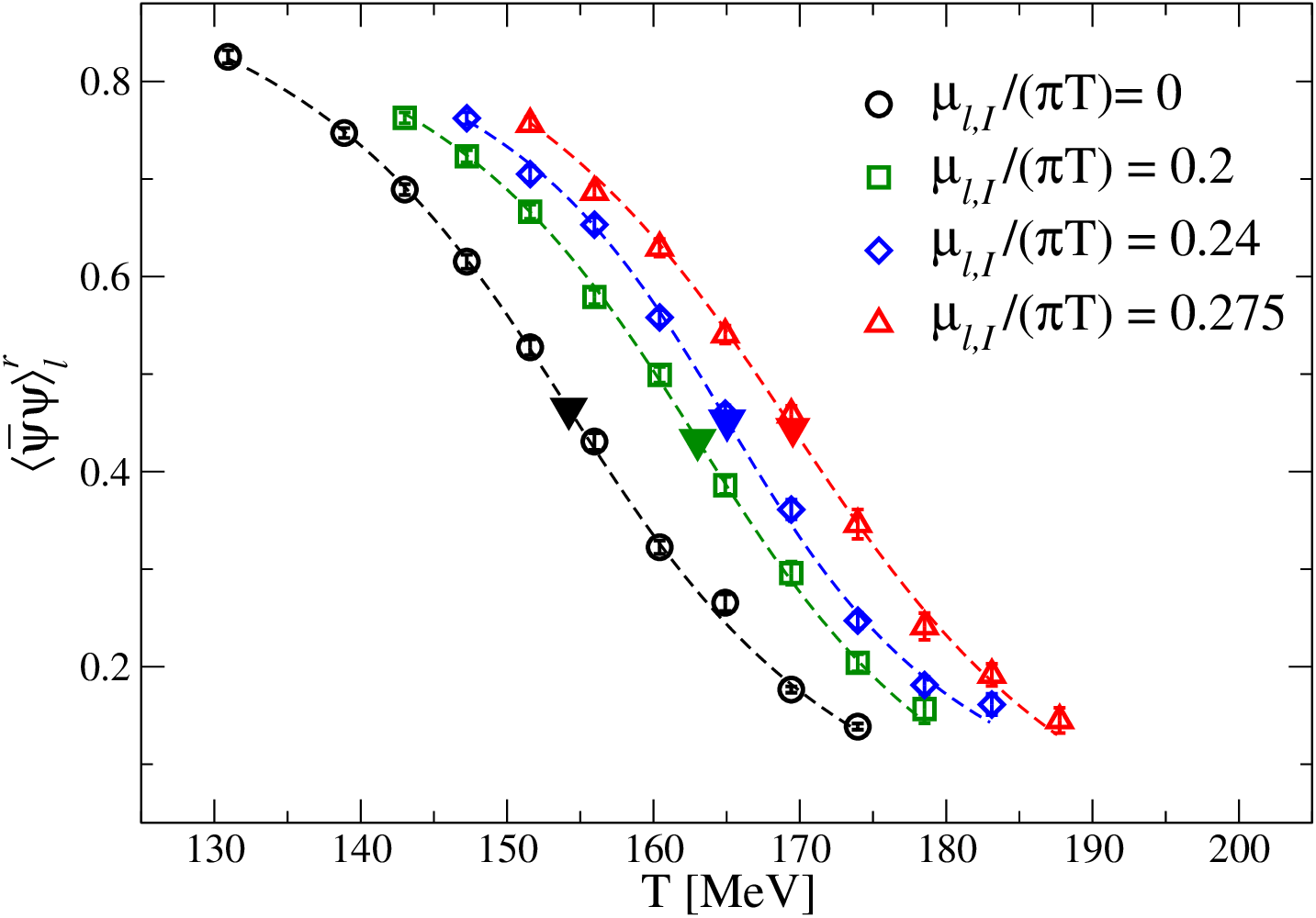}
\includegraphics[width=0.92\columnwidth, clip]{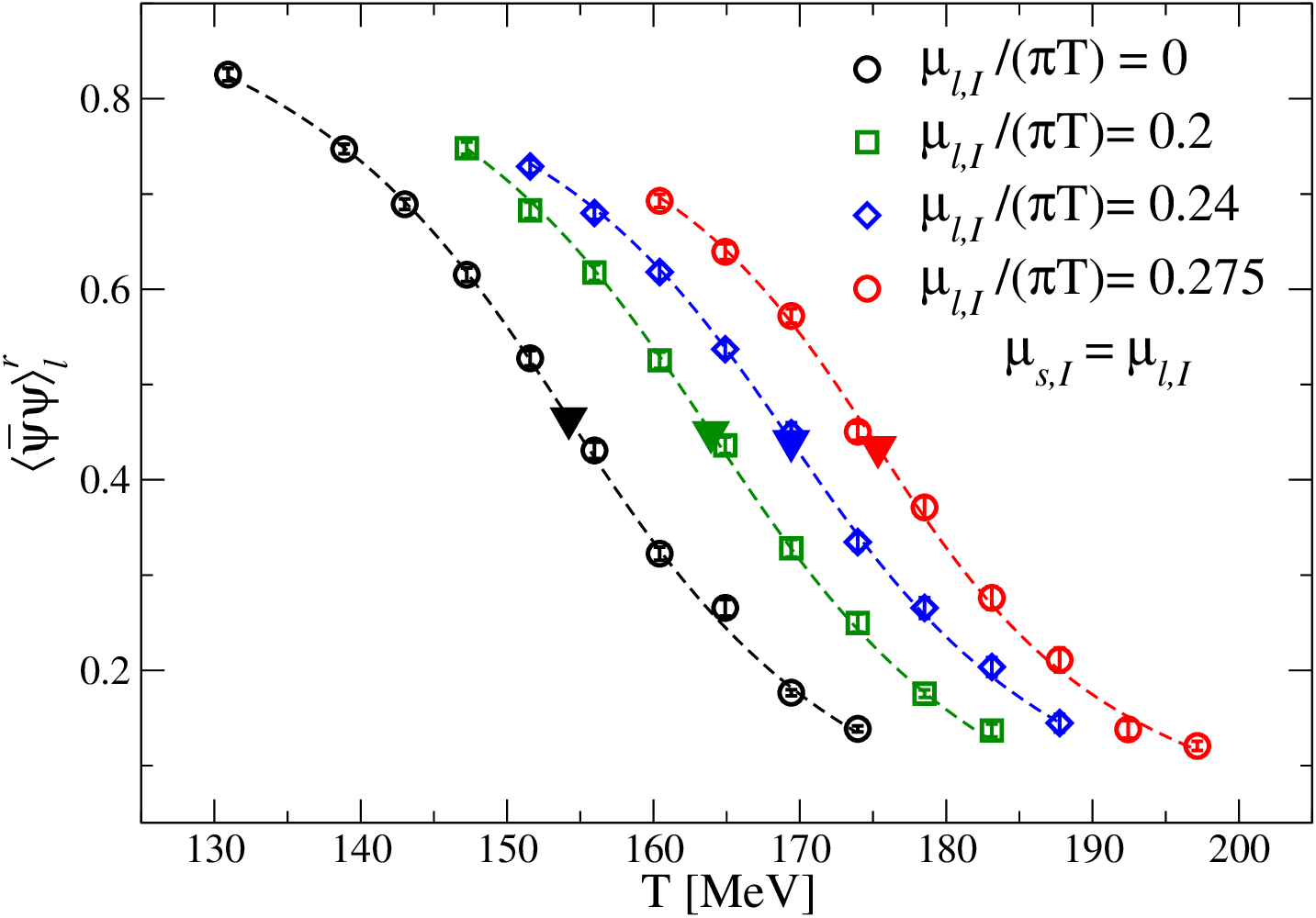}
\caption{Renormalized light chiral condensate for 
various values of $T$ and $\mu_{l}$ respectively on:
$24^3 \times 6$ lattice with $\mu_s = 0$ (top); 
$32^3 \times 8$ lattice with $\mu_s = 0$ (middle); 
$32^3 \times 8$ lattice with $\mu_s = \mu_l$ (bottom).
In the top figure the data from the $32^3\times6$ lattice at $\mu_l=0$ are also
shown for comparison. {Lines correspond to the best fit described in the text
and the filled triangles denote the values at the pseudo-critical temperature.}}
\label{chircond2406and1606_mu0}
\end{figure}
\begin{figure}[h!]
\includegraphics[width=0.92\columnwidth, clip]{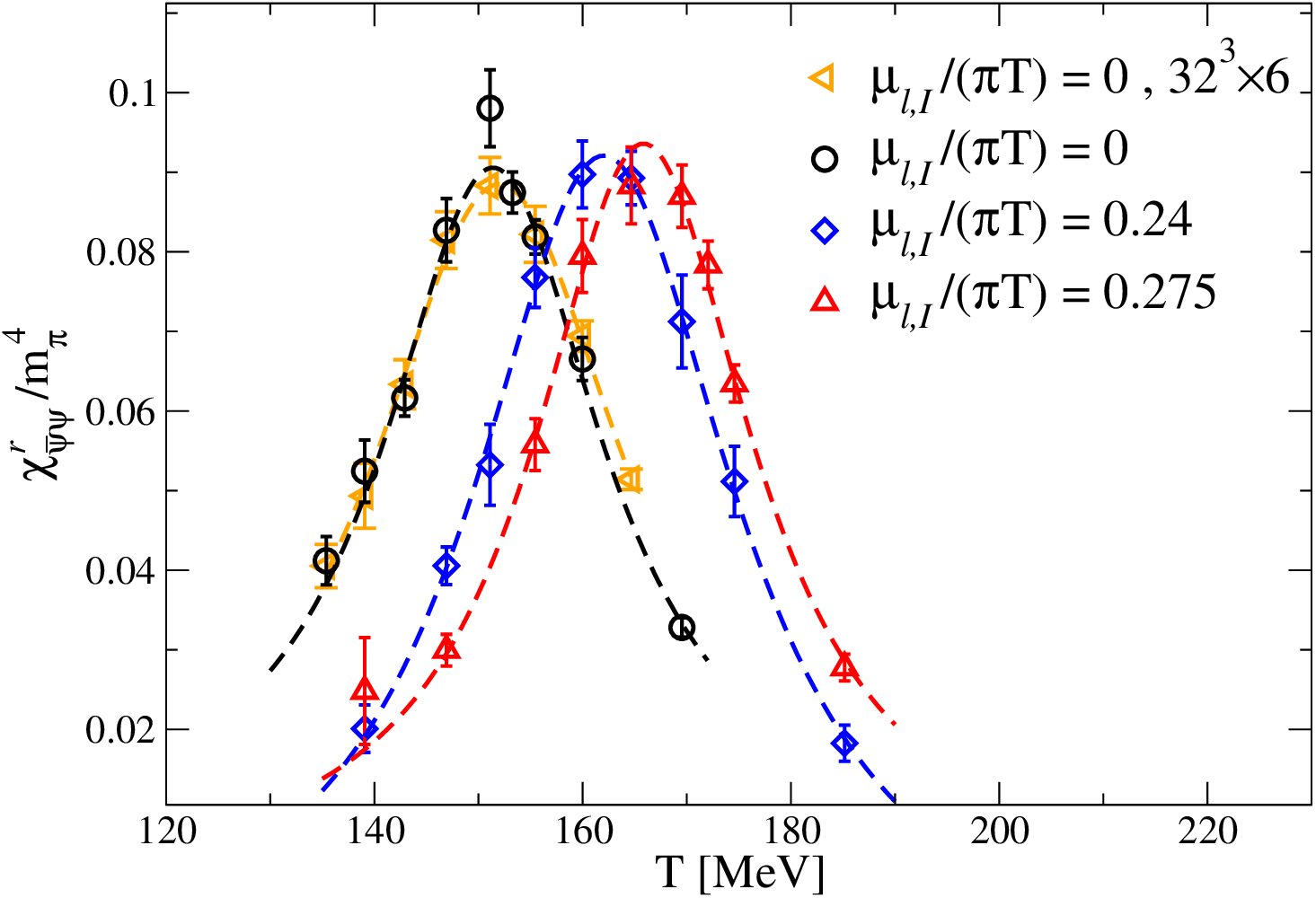}
\includegraphics[width=0.92\columnwidth, clip]{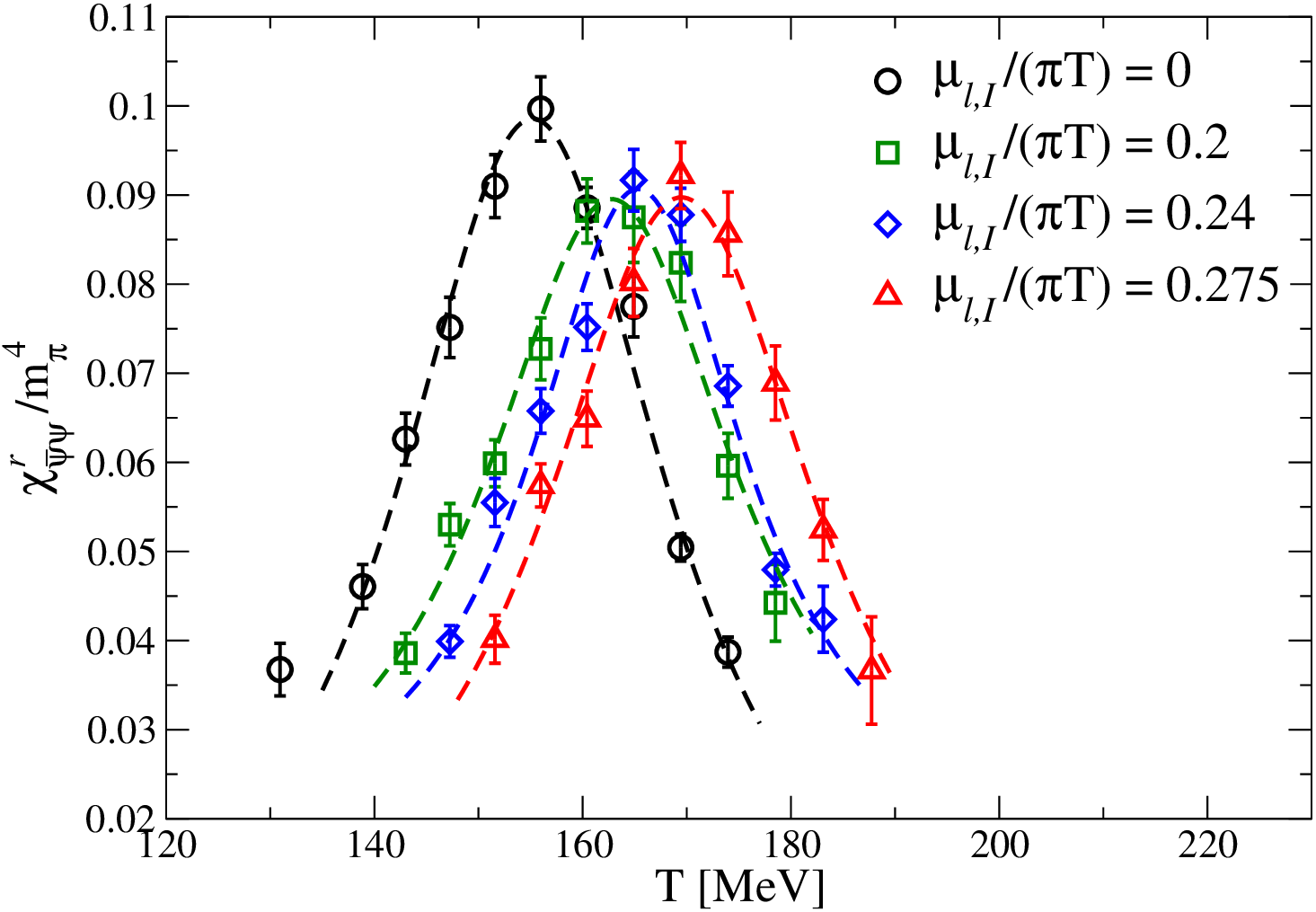}
\includegraphics[width=0.92\columnwidth, clip]{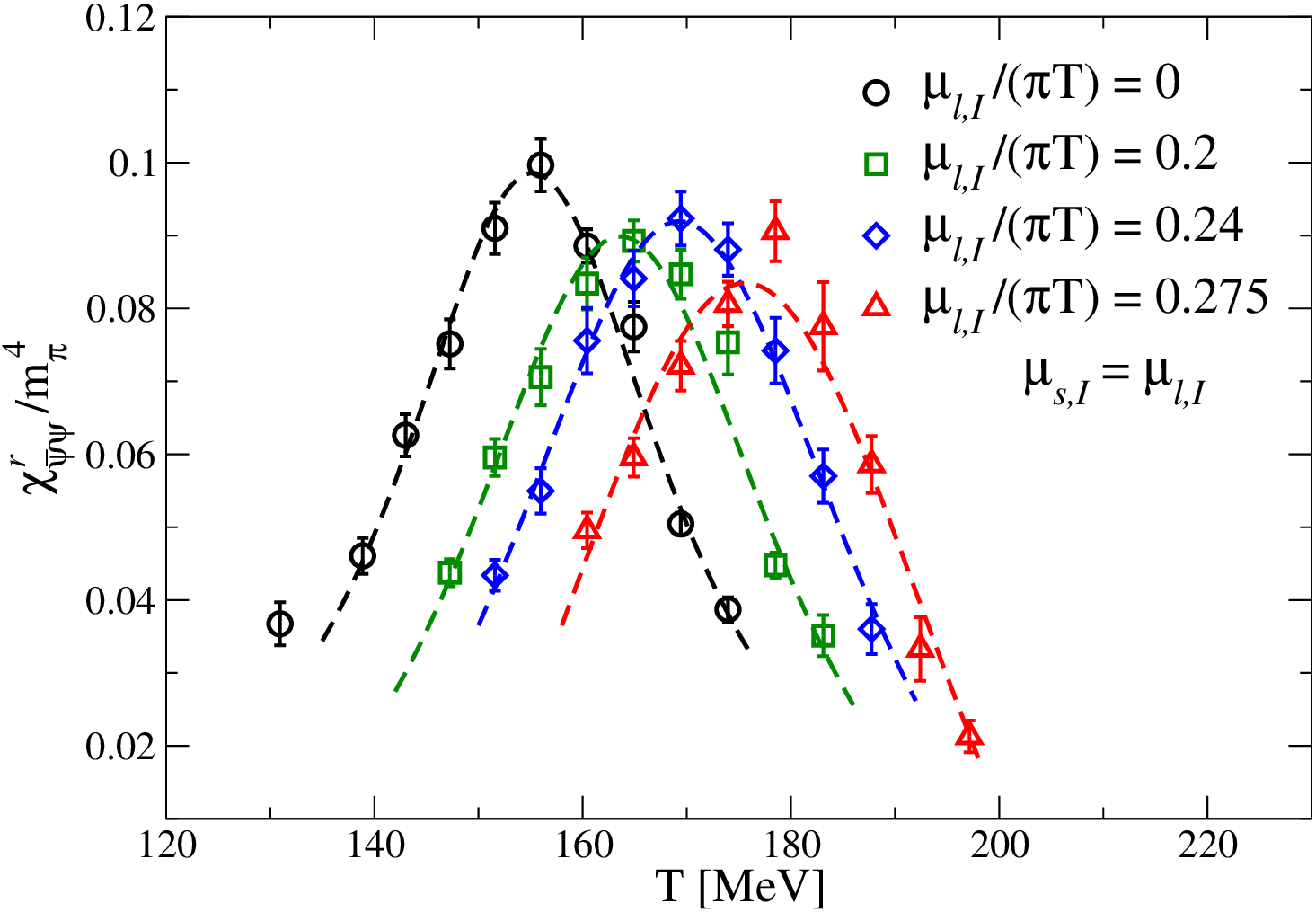}
\caption{Renormalized light chiral susceptibility, divided
by $m_\pi^4$, 
for various values of $T$ and $\mu_{l}$ respectively on:
$24^3 \times 6$ lattice with $\mu_s = 0$ (top); 
$32^3 \times 8$ lattice with $\mu_s = 0$ (middle); 
$32^3 \times 8$ lattice with $\mu_s = \mu_l$ (bottom).
In the top figure the data from the $32^3\times6$ lattice at $\mu_l=0$ are also
shown for comparison.}
\label{chirsusc2406and1606mu0}
\end{figure}

In Figs.~\ref{chircond2406and1606_mu0} and \ref{chirsusc2406and1606mu0} we plot
the results obtained respectively for the renormalized light chiral condensate
$\langle \bar{\psi}\psi\rangle^r_{l}$ and for the renormalized chiral
susceptibility $\chi^r_{\bar{\psi}\psi}$, which are our reference observables
and have been defined in Eqs.~\eqref{rencond}-\eqref{rensusc}.
Since no real phase transition is present {in the explored range of chemical potentials},
before going on we have to define {a prescription to locate the} 
pseudo-critical
temperature $T_c$ (a comparison with the results obtained by other definitions
and/or approaches is reported in the next section). We will adopt the two
following definitions of $T_c$, related to the two different observables
studied:
\begin{enumerate} 
\item the temperature corresponding to the inflection point of the renormalized
chiral condensate (as defined by Eq.~\eqref{rencond}) ; 
\item the temperature corresponding to the maximum of the renormalized chiral
susceptibility (as defined by Eq.~\eqref{rensusc}).  
\end{enumerate}
Both these definitions are faithful, \ie when a real phase transition is
present its location is correctly identified (in the thermodynamical limit) by
means of them. 

In order to determine the inflection point of the renormalized chiral condensate,
we performed a best fit on the data by using the expression
\begin{equation}
\label{arctan}
\langle \bar{\psi}\psi\rangle ^{r}_{l} (T) = A_1  + B_1 \arctan\left(C_1 (T-T_{c})\right)\, ,
\end{equation}
with the four independent parameters $A_1$, $B_1$, $C_1$ and $T_c$. This function is
found to well describe the behavior of $\langle \bar{\psi}\psi\rangle ^{r}_{l}
(T)$ in the whole range of explored temperatures.  The best fits obtained by
this procedure are plotted, together with the corresponding data points, in
Fig.~\ref{chircond2406and1606_mu0}, the position of the inflection point being
denoted, for each data set, by a filled triangle.  The errors on the fit
parameters have been estimated by means of a bootstrap analysis; results for
$T_c$ are stable, within the quoted errors, if a different interpolation (e.g.,
through an hyperbolic tangent) is adopted to locate the inflection point. 

In the case of the renormalized chiral susceptibility, 
{a reasonable description of the data aroud the peak location 
is provided by a Lorentzian function}
\begin{equation}\label{parabola}
\chi^r_{\bar{\psi}\psi} = \frac{A_2}{B_2^2 + (T-T_{c})^2}\, .
\end{equation}
{A fit to this function provides $T_c$, whose statistical error is again estimated by a bootstrap
analysis.  Finally, a systematic error is also taken into account and estimated
by changing, for each data set, the fitted range of temperatures around the
peak.  In Fig.~\ref{chirsusc2406and1606mu0} we report numerical data for the
dimensionless ratio $\chi^r_{\bar{\psi}\psi}/m_\pi^4$ as a function of $T$,
together with some of the fits performed}~\footnote{{Notice that if 
$T$-dependent dimensionless combinations of the 
susceptibility are adopted, like, e.g., 
$\chi_{\bar\psi \psi}^r/T^4$, the behavior deviates significantly from a Lorentzian function. Moreover, the 
locations of the maxima move to lower values of $T$ by about 5 MeV.}}.

\begin{table}[t!]
\centering
\begin{tabular}{ |c|c|c|c|c| }
\hline
Lattice & $\mu_{l,I}/(\pi T)$ & $\mu_{s,I}/(\pi T)$ & $T_c(\bar{\psi}\psi)$ &  $T_c(\chi^r)$ \\
\hline
%\rule{0mm}{3.3mm}
$16^3\times 6$ & 0.00  & 0.00   &  148.2(3)   &  150.7(4) \\ 
$16^3\times 6$ & 0.20  & 0.00   &  155.0(4)   &  157.0(4)  \\ 
$16^3\times 6$ & 0.24  & 0.00   &  158.9(4)   &  160.0(4)  \\ 
$16^3\times 6$ & 0.275 & 0.00   &  161.2(4)   &  162.7(4)  \\ 
\hline                                                            
%\rule{0mm}{3.3mm}
$24^3\times 6$ & 0.00  & 0.00   &  149.0(6)   &  151.6(5) \\ 
$24^3\times 6$ & 0.24  & 0.00   &  160.8(7)   &  162.0(5)  \\ 
$24^3\times 6$ & 0.275 & 0.00   &  164.1(4)   &  165.9(4)  \\ 
\hline                                                               
%\rule{0mm}{3.3mm}
$32^3\times 6$ & 0.00  & 0.00   &  149.1(7)   &  152.0(4) \\ 
$32^3\times 6$ & 0.24  & 0.00   &  160.2(3)   &  162.7(4)  \\ 
$32^3\times 6$ & 0.275 & 0.00   &  163.4(3)   &  165.5(4)  \\ 
\hline                                                               
%\rule{0mm}{3.3mm}
$32^3\times 8$ & 0.00  & 0.00   &  154.2(4)   &  155.6(7)\\
$32^3\times 8$ & 0.20  & 0.00   &  162.9(8)   &  163.0(6)  \\
$32^3\times 8$ & 0.24  & 0.00   &  165.0(5)   &  165.8(8)  \\
$32^3\times 8$ & 0.275 & 0.00   &  169.5(9)   &  169.8(7)  \\
$32^3\times 8$ & 0.20  & 0.20   &  163.9(6)   &  165.3(9)  \\
$32^3\times 8$ & 0.24  & 0.24   &  169.4(7)   &  169.6(7)  \\
$32^3\times 8$ & 0.275 & 0.275  &  175.4(6)   &  177.0(8)  \\
\hline
\end{tabular}
\caption{Critical temperatures obtained from the renormalized chiral condensate
and from the renormalized chiral susceptibility. Reported errors do not take
into account the uncertainty on the determination of the physical scale, which
is of the order of 2-3\,\%~\cite{tcwup1,befjkkrs}.} \label{tab:tc_all}
\end{table}

The full set of determinations of $T_{c}(\mu_{l,I}, \mu_{s,I})$ is reported in
Table~\ref{tab:tc_all}. We stress that such values do not take into account the
error on the determination of the physical scale, which is of the order of
2-3\,\%~\cite{tcwup1,befjkkrs}; on the other hand, since such error affects all
$T_c$ values in the same way, its effect on the ratio of pseudo-critical
temperatures, which is the quantity entering the determination of $\kappa$, is
expected to be negligible.

\begin{figure}[h]
\includegraphics[width=0.92\columnwidth, clip]{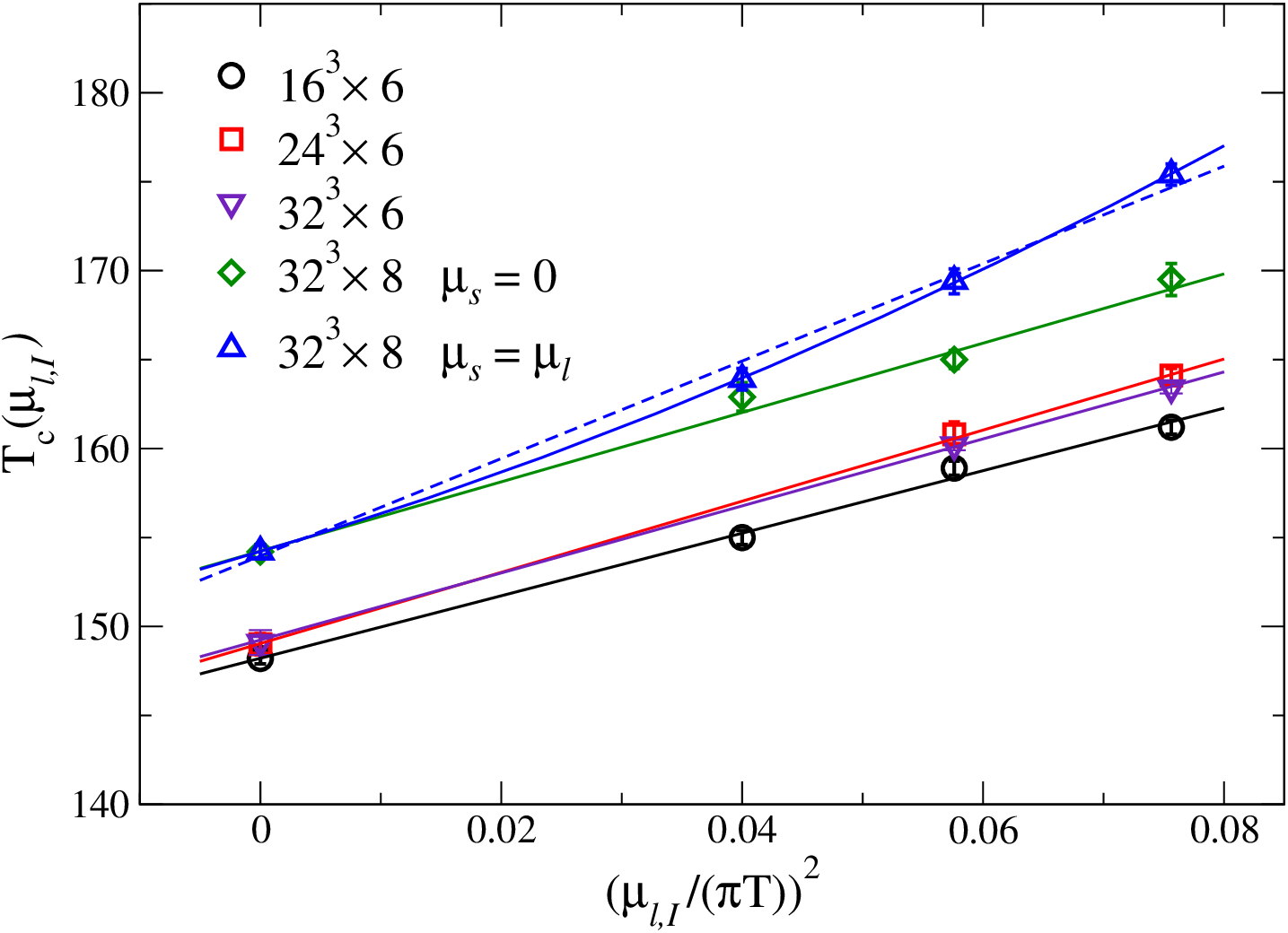}
\caption{Determinations of $T_{c}$ obtained from the renormalized chiral
condensate $\langle \bar{\psi}\psi\rangle^r_l$ for various values of the chemical
potential and lattice sizes.  The lines correspond to quadratic and quartic
fits in $\mu_{l,I}$, as discussed in the text.  Fit results are reported in
Table \ref{tab:fit_cond}.}
\label{tcvsmu_cond}
\end{figure}

\begin{figure}[h]
\includegraphics[width=0.92\columnwidth, clip]{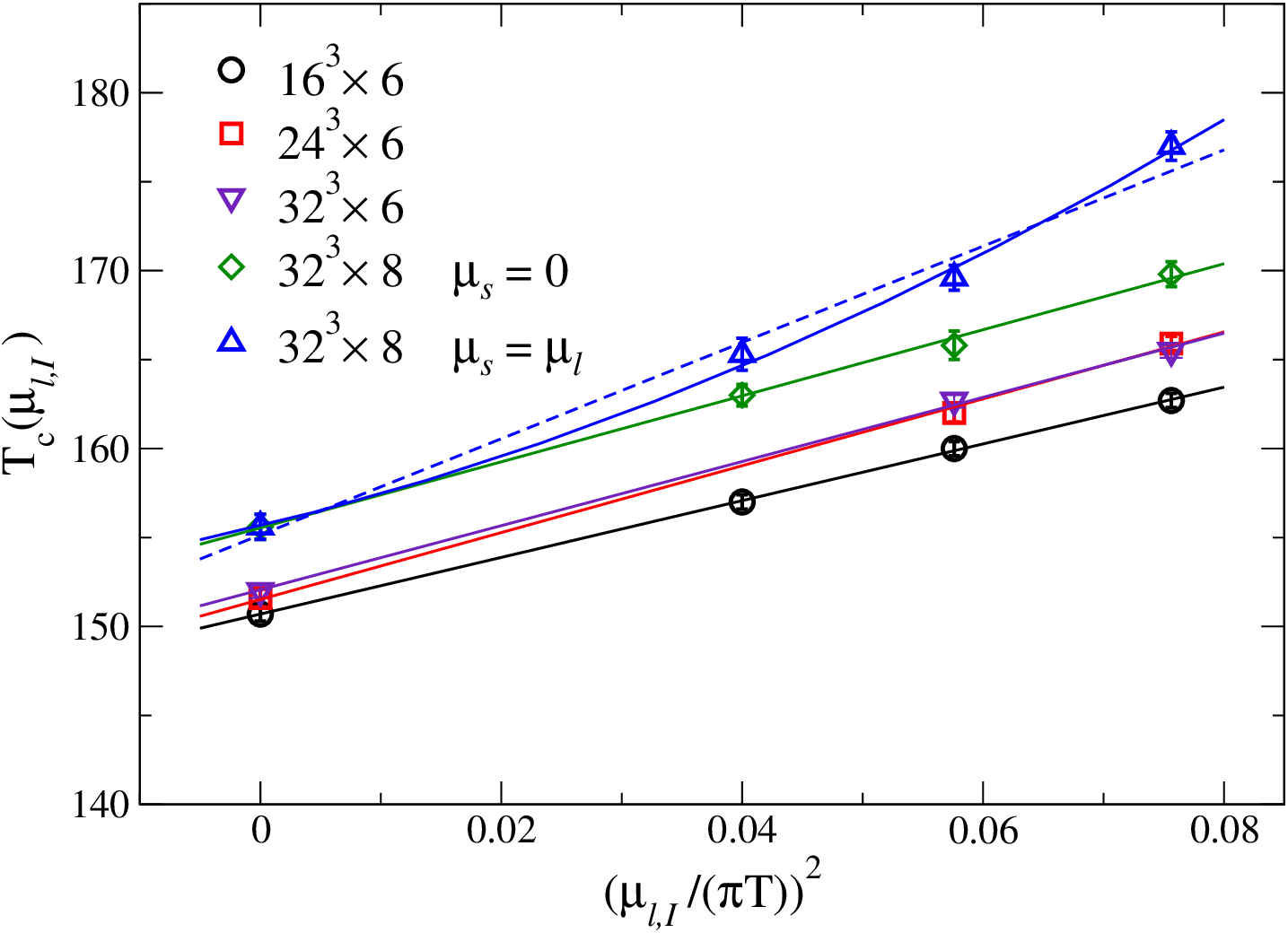}
\caption{Determinations of $T_{c}$ obtained from the renormalized chiral
susceptibility $\chi_{\bar{\psi}\psi}^r$ for various values of the chemical
potential and lattice sizes.  The lines correspond to quadratic and quartic
fits in $\mu_{l,I}$, as discussed in the text.  Fit results are reported in
Table~\ref{tab:fit_susc}.}
\label{tcvsmu_susc}
\end{figure}

In order to extract the curvature, we performed a fit to the values obtained
for $T_c(\mu_{l,I})$, separately for each lattice size and setup of chemical
potentials, according to the function\footnote{{Both $\kappa$ and $b$ are 
normalized as the coefficients that would appear in the expansion in 
terms of $\mu_B$; this is the reason of the factors 9 and 81 appearing 
in Eq.~(\ref{fitfun}). Notice also that, going to the fourth
order expansion, one needs to specify what is the temperature
appearing in the ratio $\mu/T$, as we have done in Eq.~(\ref{fitfun}).
}}
\begin{equation}
\begin{aligned}
T_c(\mu_{l,I}) &= T_c(0)\, \left( 1 + 
9 \kappa \left( \frac{\mu_{l,I}}{T_c(\mu_{l,I})} \right)^2 + \right.\\
&+ \left. 81\, b \left( \frac{\mu_{l,I}}{T_c(\mu_{l,I})} \right)^4 + O(\mu_{l,I}^6) \right) \, .
\label{fitfun}
\end{aligned}
\end{equation}
In this way we got estimates of $\kappa$ for all the lattices and the chemical
potential setups adopted.  The results of these fits are reported in
Tables~\ref{tab:fit_cond} and \ref{tab:fit_susc}, for the critical temperatures
obtained  from the chiral condensate and for the chiral susceptibility
respectively.  In Figs.~\ref{tcvsmu_cond} and \ref{tcvsmu_susc} data for
$T_c(\mu_{l,I})$ are plotted together with the results of the aforementioned
fits.  In most cases, a simple linear fit (i.e. setting $b = 0$) works quite
well; just for the $\mu_{s}=\mu_l$ setup (studied only on the $32^3\times 8$
lattice) the introduction of a quartic correction is necessary in order to
obtain reasonable values of the $\tilde \chi^2$ test.  It is tempting to
associate the enhancement of non-linear corrections in the $\mu_s = \mu_l$
setup to the fact that, in this case, the Roberge-Weiss endpoint is closer to
the $\mu_l = 0$ axis, so that the associated critical behavior might have a
stronger influence on the small $\mu_l$ region.

%%%%%%%%%%%%%%%%%%%%%% TABELLE %%%%%%%%%%%%%%%%%%%%%%%%%%%%%%%%%%%%%%%%%
\begin{table}[h]
\centering
\begin{tabular}{|c|c|c|c|c|c|}
\hline
Lattice & $\mu_s$ & $T_c(0)$ & $\kappa$ & $b$ & $\chi^2/n_{dof}$ \\
\hline
$16^3\times 6$ & 0.00   & 148.2(3) & 0.0133(4)  & - & 1.4 \\
$24^3\times 6$ & 0.00   & 149.1(6) & 0.0150(7)  & - & 0.17 \\
$32^3\times 6$ & 0.00   & 149.2(7) & 0.0142(8)  & - & 0.2 \\
$32^3\times 8$ & 0.00   & 154.2(4) & 0.0142(7)  & - & 1.2 \\
$32^3\times 8$ & $\mu_l$& 154.0(4) & 0.0200(6)  & - & 2.5 \\
$32^3\times 8$ & $\mu_l$& 154.2(4) & 0.0149(24)  & 0.0008(4) & 0.04 \\
\hline
\end{tabular}
\caption{Parameters of the best fit to $T_c(\mu_{l,I})$ 
from the renormalized chiral
condensate according to Eq.~(\ref{fitfun}).  Blank fields indicate that the
corresponding parameter has been set to zero in that fit.}
\label{tab:fit_cond}
\end{table}

\begin{table}[h]
\centering
\begin{tabular}{ |c|c|c|c|c|c| }
\hline
Lattice & $\mu_s$ & $T_c(0)$ & $\kappa$ & $b$ & $\chi^2/n_{dof}$ \\
\hline
$16^3\times 6$ & 0.00   & 150.7(4) & 0.0119(6)  & - & 0.1 \\
$24^3\times 6$ & 0.00   & 151.5(5) & 0.0140(7)  & - & 0.7 \\
$32^3\times 6$ & 0.00   & 152.1(3) & 0.0134(5)  & - & 0.4 \\
$32^3\times 8$ & 0.00   & 155.6(6) & 0.0134(9)  & - & 0.2 \\
$32^3\times 8$ & $\mu_l$   & 155.2(6) & 0.0196(10)  & - & 3.3 \\
$32^3\times 8$ & $\mu_l$   & 155.6(7) & 0.012(3)  & 0.0010(5) & 1.2 \\
\hline
\end{tabular}
\caption{The same as in Table~\ref{tab:fit_cond}, but using the critical
temperatures estimated from the maxima of the renormalized chiral
susceptibility.}
\label{tab:fit_susc}
\end{table}

%%%%%%%%%%%%%%%%%%%%%%%%%%%%%%%%%%%%%%%%%%%%%%%%%%%%%%%%%%%%5

\begin{figure}[h]
\includegraphics[width=0.92\columnwidth, clip]{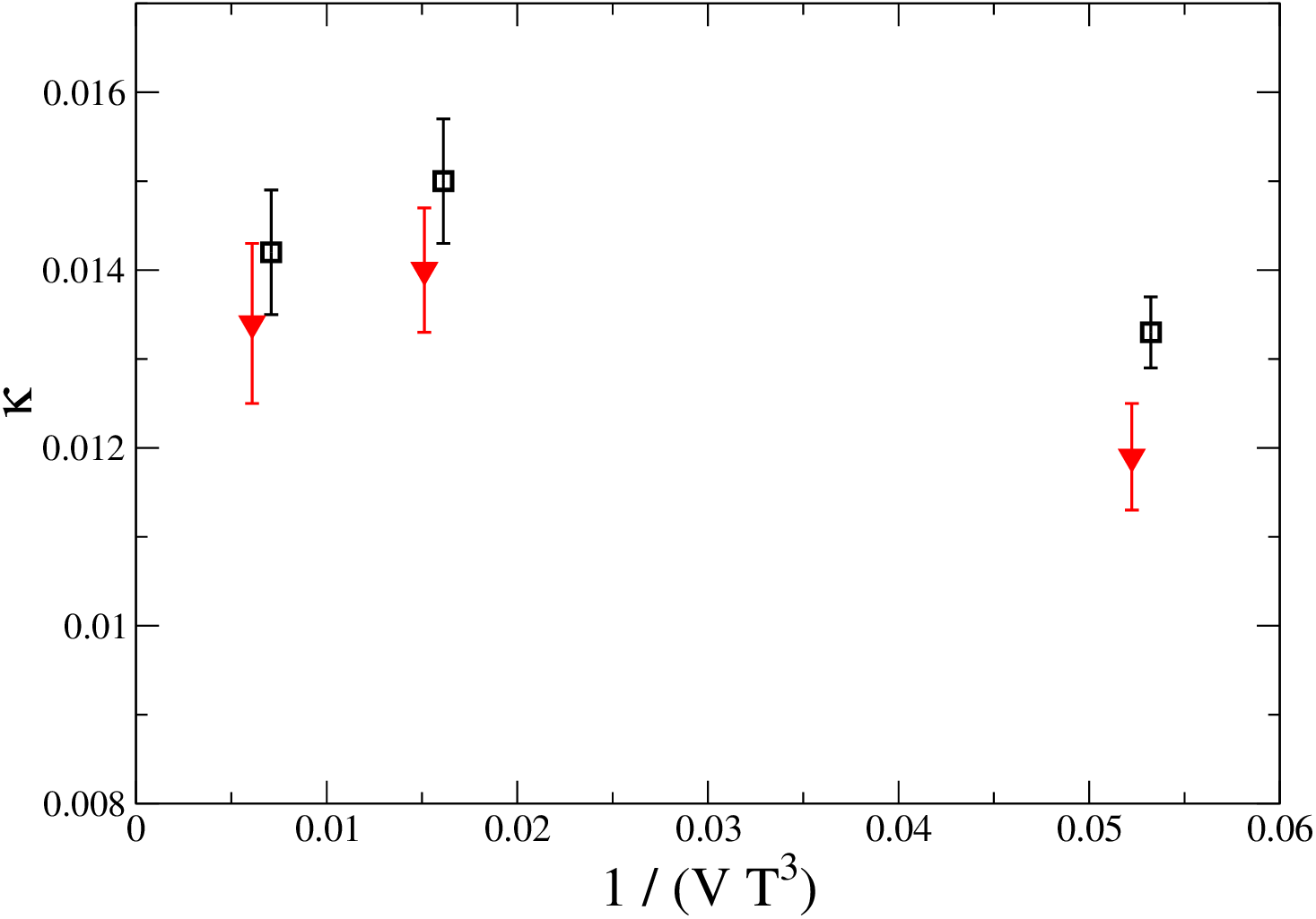}
\caption{Fitted values of the curvature $\kappa$ from the $N_t =6$
lattices as a function of the inverse spatial volume.
Squares correspond to the determinations obtained by
using the chiral condensate, triangles to the chiral susceptibility estimates.}
\label{finitesize}
\end{figure}

\subsection{Discussion of results}

Let us now analyze the main features emerging from our results.  A
first important point is that, as one would expect, the value of $T_c$ at zero
chemical potential is in agreement with other existing determinations in the
literature~\cite{aefks,afks,betal,tchot,tchot2}, i.e. in a range around 155
MeV.

{
Comparing data at the same lattice spacing and different spatial volumes 
($N_t = 6$ and $L_s = 16, 24, 32$)}, or at the same {physical} spatial 
volume\footnote{{Lattices with the same aspect ratio $L_s/N_t$
corresponds to approximately equal spatial volumes at the crossover,
apart from the residual $a$-dependence of $T_c$.}}
and different lattice spacings ($24^3 \times 6$ vs $32^3 \times 8$), one
concludes that both finite size and finite lattice spacing effects are visible
in the determination of the pseudo-critical temperature at zero and nonzero
$\mu_l$, both tending in general to decrease the value of $T_c$.  It is also
evident that the introduction of a non-zero $\mu_{s,I} = \mu_{l,I}$ has a
significant impact, leading to a relative temperature change
$T_c(\mu_{l,I})/T_c(0) - 1$ which is up to 40\,\% larger {(at the
largest value of $\mu_{l,I}$ explored)}, with respect to the
$\mu_{s,I} = 0$ case.

On the other hand, when the dependence of $T_c$ on $\mu_{l,I}$ is considered,
in order to extract the curvature, good part of these effects boils down to a
constant shift of the curves or to the introduction of quartic corrections (see
Figs.~\ref{tcvsmu_cond} and \ref{tcvsmu_susc}).  That means, in particular,
that the curvature $\kappa$ is a more stable quantity: the introduction of the
strange quark chemical potential does not modify it within present errors,
finite lattice spacing effects seem to be within the 10\,\%
level. {Finite size effects are of the order of 15\,\% when going from lattices
with aspect ratio $\sim$2.7 to lattices with aspect ratio 4. However they are 
much smaller and stay within statistical errors when going from aspect ratio 4 to aspect
ratio $\sim$5.3, suggesting that they are well under control already on lattices
with aspect ratio 4; all that can be 
appreciated from Fig.~\ref{finitesize}, where we report
our determinations of $\kappa$ for $N_t = 6$ and different spatial volumes.} 

\begin{figure}[h!]
\includegraphics[width=0.92\columnwidth, clip]{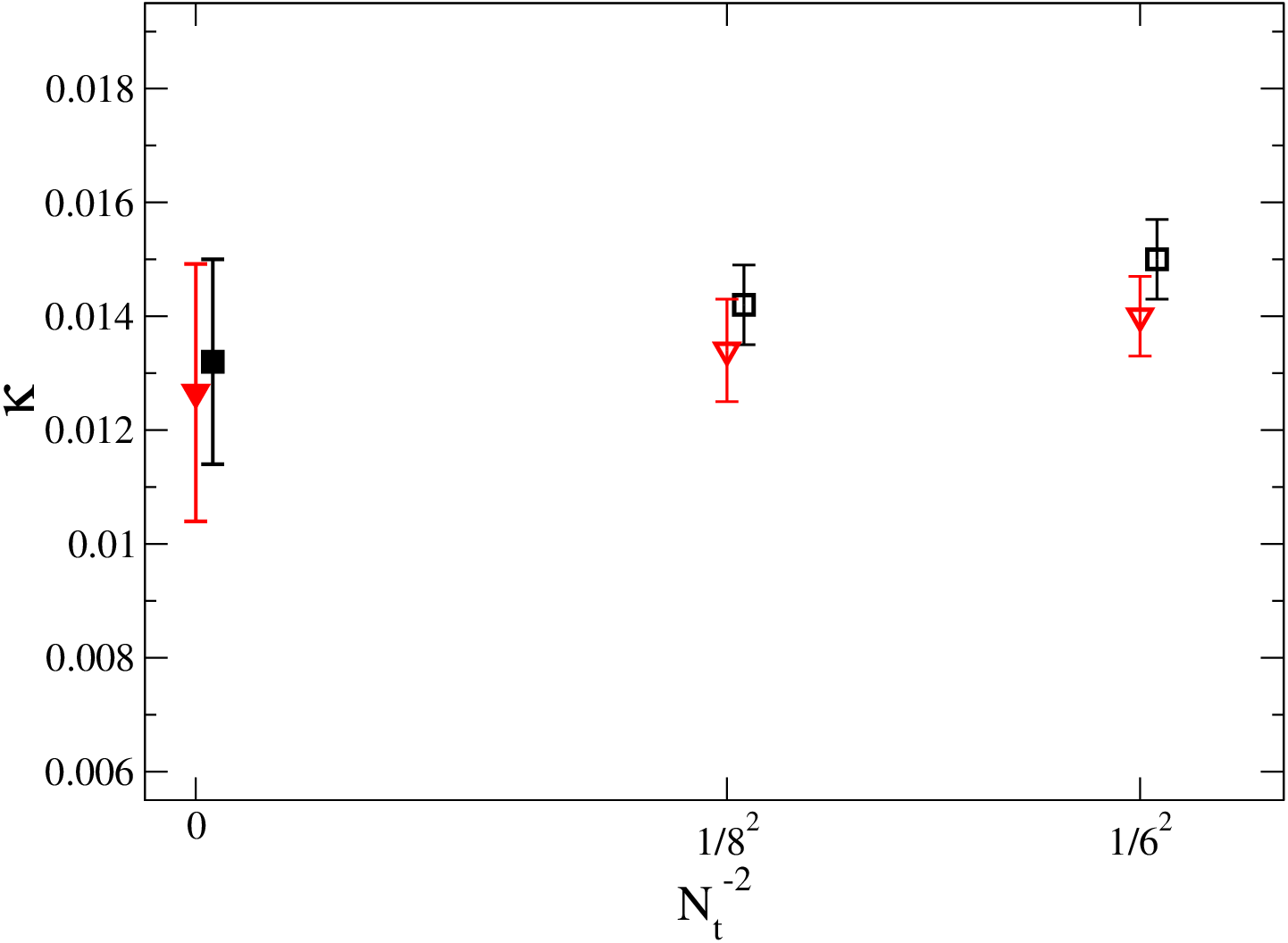}
\caption{Determinations of the critical line curvature $\kappa$ as a function
of $N_t^{-2}$. {Squares correspond to the determinations obtained by
using the chiral condensate, triangles to the chiral susceptibility estimates.
The points at zero abscissa correspond to a rudimentary continuum limit
extrapolation, assuming corrections linear in $N_t^{-2}$.}}
\label{kappa}
\end{figure}

In Fig. \ref{kappa} we have reported our determinations of $\kappa$ obtained
for $\mu_{s,I} = 0$ and on the lattices with aspect ratio 4, together with a
rudimentary continuum extrapolation, performed assuming order $a^2$
corrections. This cannot be considered as a rigorous continuum extrapolation
yet, since we have not enough data points to perform a best fit. The numbers we
obtain are $\kappa = 0.0132(18)$ from the renormalized chiral condensate and
$\kappa = 0.0126(22)$ from the renormalized chiral susceptibility, {hence we have a very good
agreement between the two determinations.}

\section{Comparison with other determinations}
\label{sec5}

\begin{figure}[h!]
\includegraphics[width=0.9\columnwidth, clip]{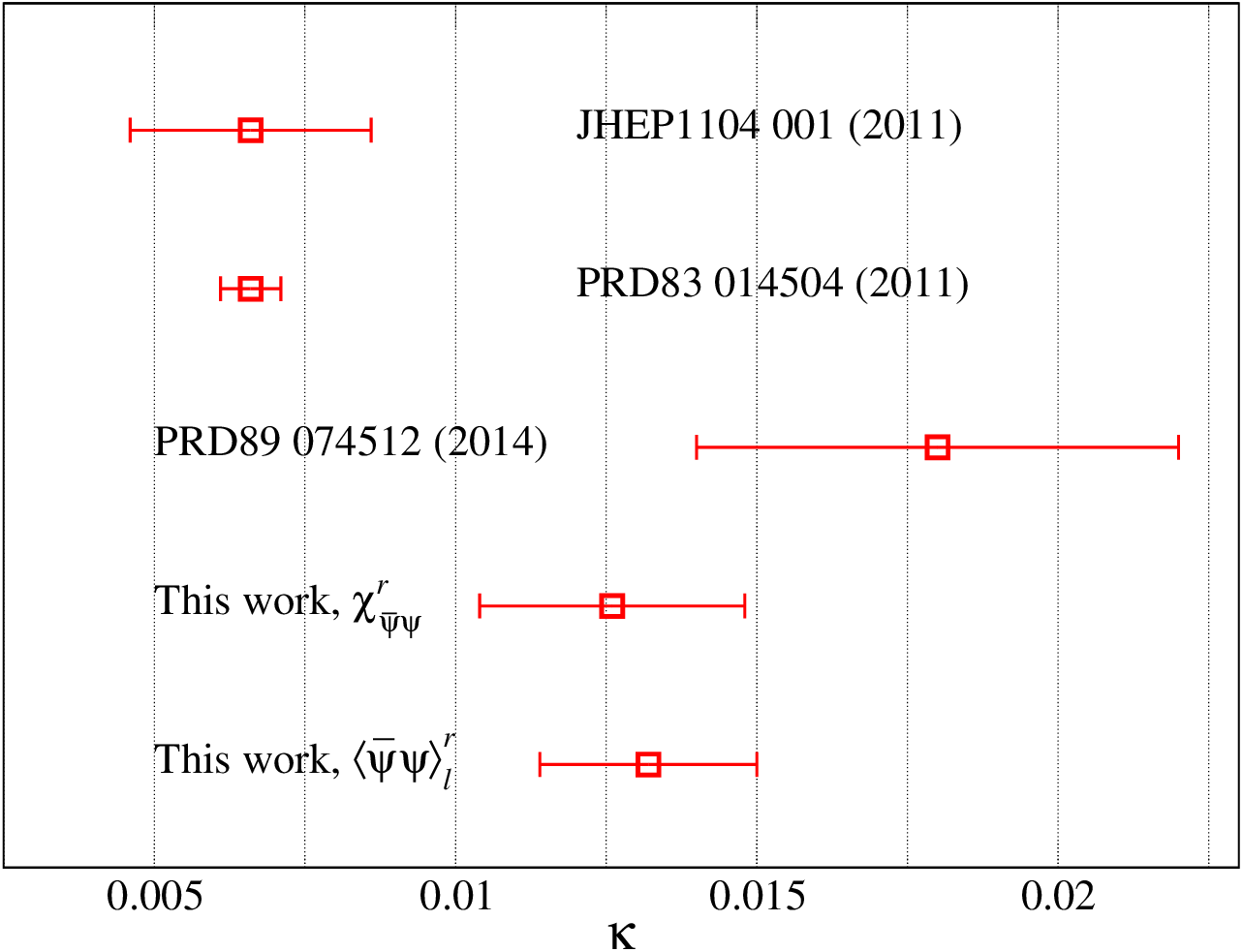}
\caption{Determinations of the critical line curvature $\kappa$ in different works. From bottom to top: 
 \emph{i}) Analytic continuation, renormalized chiral condensate, this work;
 \emph{ii}) Analytic continuation, renormalized chiral suscepbtibility, this work;
 \emph{iii}) Analytic continuation, disconnected chiral susceptibility with $\mu_s = \mu_l$, Ref.~\cite{ccp}; 
 \emph{iv}) Taylor expansion, chiral susceptibility, Ref.~\cite{Kaczmarek2011}; 
 \emph{v}) Taylor expansion, chiral condensate (different renormalization), Ref.~\cite{Endrodi2011}.}
\label{k_literature}
\end{figure}

In Fig.~\ref{k_literature} we compare our present results with previous
ones in the literature. We do not report many early
determinations and consider only a
collection of recent ones, {which look at the chiral transition and} 
have been obtained by discretizations
of $N_f = 2+1$ 
QCD at or close  to the physical point~\cite{Kaczmarek2011,
Endrodi2011, ccp}.  Our results seem generally larger than results obtained by
the Taylor expansion~\cite{Kaczmarek2011, Endrodi2011} and in {marginal} agreeement
with results obtained by analytic continuation and a different
discretization~\cite{ccp}.  However, the correct assessment of possible
discrepancies or agreement requires a careful analysis of the possible sources
of systematic differences between the various determinations, a task that we
try to accomplish in this Section.

Part of the effects are related to the different lattice discretizations
adopted and should disappear as one approaches the continuum limit.
Moreover, since no real transition takes place, the result depends
on the particular physical 
quantity and on the
prescription chosen to locate the crossover at zero and non-zero $\mu_B$. In
this work, we have considered either the chiral condensate (renormalized as in
Eq.~(\ref{rencond})) and its inflection point, or the full chiral susceptibility
(renormalized as in Eq.~(\ref{rensusc})) and its maximum: as we have already
discussed, both are faithful, in the sense that provide a correct location of
$T_c$ in the case of a real transition.

The determination in Ref.~\cite{Kaczmarek2011} develops on previous studies {carried on} by
the same group {about} the chiral transition at $m_{l}=0$ with physical 
$m_s$~\cite{Ejiri2009}. The basic idea is that, if 
{for physical $m_s$} the chiral transition is second order, 
the neighborhood of the critical point can be described by two scaling
variables, $t$ and $h$. 
To leading order only $h$ depends on the chiral
symmetry breaking parameter, \ie $m_l$, and we thus have the relations
\begin{equation}\label{eq:t_h}
t \simeq \frac{1}{t_0}\left(\frac{T-T_c(0)}{T_c(0)}
+\kappa\left(\frac{\mu_B}{T_c(0)}\right)^2\right) \quad
h \simeq \frac{1}{h_0}\frac{m_{l}}{m_s}\ ,
\end{equation} 
where $t_0$ and $h_0$ are dimensionless factors. These can be fixed by imposing
{appropriate} normalization conditions to the scaling functions (see \cite{Ejiri2009,
Kaczmarek2011} for more details). 

In Eq.~\eqref{eq:t_h} we denoted by $T_c(0)$ the critical temperature at
vanishing chemical potential and, since the transition for generic $\mu_B$ is
located at $t=0$, it follows that $\kappa$ is the curvature of the critical
line as previously defined in Eq.~\eqref{corcur}. 
To extract the value of $\kappa$ one can study an observable $\varphi$ 
directly related to the critical behavior, like the chiral condensate,
which plays {a} role {analogous} to the magnetization and, in the scaling
regime, is governed by a well known scaling behavior
$\varphi\equiv\varphi(t,h)$, which is fixed according to the $O(4)$ 
universality class and {was} checked at $\mu_B = 0$
in Ref.~\cite{Ejiri2009}. It is easy then to prove, by means of Eq.~(\ref{eq:t_h}), that
\begin{equation}\label{eq:k_hotqcd}
\kappa=\frac{t_0}{\partial_t \varphi} \frac{\partial\varphi}{\partial (\mu_B/T)^2} \,.
\end{equation}
{In Ref.~\cite{Kaczmarek2011}}, ${\partial\varphi}/{\partial
(\mu_B/T)^2}$ {was} measured directly in terms of a mixed
susceptibility computed at $\mu_B = 0$, while ${t_0}/{\partial_t \varphi}$
{was} fixed by the $O(4)$ scaling function.  In this way the
value of $\kappa$ {was} inferred by imposing a scaling
behavior for the mixed susceptibility computed for different values of the
light quark mass. 

In Ref.~\cite{Laermann2013} a variant of {this approach} was proposed, which makes use
of simulations performed at imaginary chemical potential (with
$\mu_{l}=\mu_s$). Having at disposal data obtained at $\mu_B\neq 0$,
Eq.~\eqref{eq:t_h} can be used without the need of computing derivatives of
observables: results are compatible with those of Ref.~\cite{Kaczmarek2011}.

In this case, a direct comparison with our determination
is not easy, since one has different lattice discretizations
(p4 staggered action vs stout smeared staggered action) and no proper
continuum extrapolation from both sides 
(Ref.~\cite{Kaczmarek2011} has lattices with $N_t = 4,8$, 
Ref.~\cite{Laermann2013} has lattices with $N_t = 4$).
Moreover, one should notice that 
the value of $\kappa$ obtained in this way is 
{actually} the curvature of the second order line
in the chiral limit $m_l = 0$, {assuming $O(4)$
critical behavior;
the expectation is that
the dependence of $\kappa$ on the light quark mass is very mild.}
\\

A different procedure was put forward in Ref.~\cite{Endrodi2011}: for each
observable $\phi(T,\mu_B)$ that is monotonic in the neighborhood of the
$\mu_B=0$ transition, the authors define the critical temperature at finite
chemical potential (denoted by $T_c(\mu_B)$) as the solution of the equation
\begin{equation}\label{eq:tc_mu_wupp}
\phi(T_c(\mu_B),\mu_B)=\phi(T_c(0), 0)\,.
\end{equation}
With this definition, along the $(T_c(\mu_B), \mu_B)$ curve we have $\mathrm{d}\phi\equiv 0$, 
thus we obtain
\begin{equation}\label{eq:k_wupp}
\begin{aligned}
\kappa & \equiv-T_c(0)\left.\frac{\mathrm{d}T_c(\mu_B)}{\mathrm{d}\mu_B^2}\right|_{\mu_B=0}=\\
& =T_c(0)\left.\frac{\partial\phi/\partial\mu_B^2}{\partial\phi/\partial T}\right|_{\mu_B=0;\ T=T_c}\,.
\end{aligned}
\end{equation} 

The strangeness susceptibility and the chiral condensate were used as the $\phi$ 
observable in Ref.~\cite{Endrodi2011}, with the following renormalization prescription for 
the chiral condensate:
\begin{equation}\label{eq:ren_pres_wupp}
\langle \bar{\psi}\psi\rangle^r_{(2)}=\frac{m_{l}}{m_{\pi}^4}\left(\langle\bar{\psi}\psi\rangle_{l}
-\langle\bar{\psi}\psi\rangle_{l}(T=0)\right)\,. 
\end{equation}
{In Eq.~(\ref{eq:k_wupp}), the derivative with respect to 
$\mu_B^2$ is given in terms of a susceptibility computed 
at $\mu_B = 0$, as for Eq.~(\ref{eq:t_h}), 
while $\partial\phi/\partial T$ is obtained
directly by numerical differentiation of data at various temperatures.} 
{Notice that this
prescription for locating $T_c$ might not be faithful in the particular
case of a real transition and if the chosen observable is not an order parameter
vanishing at $T_c$: indeed, in general, the value {taken by} 
the observable at $T_c$ could change as the transition changes with $\mu_B$.}

In this case, a detailed comparison with our determination makes sense, since
we adopt the same lattice discretization and the same physical observable
(chiral condensate), even if with a different renormalization prescription.  In
particular we can understand, making use of our data, what is the influence on
the curvature of adopting a different prescription for locating $T_c$ and/or of
adopting a different renormalization prescription for the chiral condensate
(Eq.~(\ref{rencond}) vs Eq.~(\ref{eq:ren_pres_wupp})). 

{In }Fig.~\ref{chircond2406and1606_mu0} {we can see} that,
if we use the inflection points (marked with filled triangles in the plots)
as a definition of $T_c(\mu_B)$,
Eq.~\eqref{eq:tc_mu_wupp} is only
approximately satisfied, in particular at these points
the condensate assumes values  
$\langle\bar{\psi}\psi\rangle_l^r(T_c(\mu_{l,I}),\mu_{l,I})$ that
decreases {as $\mu_{l,I}$ is increased}. 
Therefore, adopting {the} prescrition {of} Ref.~\cite{Endrodi2011}
and defining $T_c(\mu_{l,I})$ as the temperature for which the condensate takes
the same value as for $T_c(0)$, we {would} obtain lower estimates of
$T_c(\mu_{l,I})$ and of $\kappa$. This is indeed what happens, as shown in Appendix~\ref{appendix:comparisons}. 
{Despite this consideration, however, the continuum extrapolated value 
of $\kappa$ does not differ within the statistical errors.

A more substantial difference is obtained if, in addition, one also adopt the
renormalization prescription of Ref.~\cite{Endrodi2011}, i.e.
Eq.~(\ref{eq:ren_pres_wupp}) (see Appendix~\ref{appendix:comparisons} for
details). In this case our continuum extrapolated value would go from $\kappa =
0.0132(18)$ to $\kappa = 0.0110(18)$, i.e. an effect of around 20\%.
Therefore, about one third of the discrepancy with {respect to} Ref.~\cite{Endrodi2011},
claiming
$\kappa = 0.0066(20)$, can be attributed to systematic effects
connected to different prescriptions for renormalization or location of $T_c$:
this is not unexpected, in view of the fact that there is no real phase
transition in the range of chemical potentials under study.
Taking that into account, the remaining discrepancy {between the two determinations} goes below 2
standard deviations. In the future, a more rigorous continuum extrapolation of
our data could better clarify the issue.
\\

{Finally, we compare with the results published in Ref.~\cite{ccp}, in which the
authors adopt the method of analytic continuation, with the setup
$\mu_{l}=\mu_s$, and locate $T_c$ by looking for the maximum of the
disconnected part of the unrenormalized chiral susceptibility. 
{The outcome of such analysis}
is in marginal agreement with our determination. One should
also take into account that, 
as illustrated in
appendix~\ref{appendix:comparisons}, adopting the {unrenormalized}
disconnected susceptibility {in place of the full renormalized one}
leads to an increased curvature: in our
case the continuum extrapolated value goes from $\kappa = 0.0126(22)$ to
$\kappa = 0.0146(41)$, hence in better agreement with the outcome of 
Ref.~\cite{ccp}.
That, taking into account that a different discretization was
used {in Ref.~\cite{ccp}} (HISQ action), {is compatible with} the
absence of significant lattice artifacts 
in both cases. 
}

Regarding the different setup of chemical potentials, 
we have already verified on our results that the introduction 
of a non-zero $\mu_s$ does not influence {the value of} $\kappa$ significantly,
but on the other hand {it} introduces larger non-linear corrections
in $\mu_{l,I}^2$. In Ref.~\cite{ccp} there was no evidence 
of such non-linear corrections, even if {only on the smallest lattice, namely $16^3 \times 6$,} 
multiple values of $\mu_{l,I}$ were explored.

\section{Conclusions}
\label{sec6}

We have presented a determination of the pseudo-critical line of $N_f = 2+1$ QCD
with physical quark masses by the method of analytic {continuation} from imaginary chemical
potentials. We considered a stout smeared staggered discretization and
performed simulations on lattices with 
$N_t = 6$ and $N_t = 8$.

In order to locate the pseudo-critical temperature $T_c$,
we have considered both the inflection point of the renormalized
condensate or the location of the peak of the renormalized chiral
susceptibility. The pseudo-critical temperature at zero quark 
chemical potential is {found to be} in agreement with previous determinations.
{Finite size effects on the curvature have been shown to be 
not significant, within present statistical errors, 
on lattices with aspect ratio 4.
A preliminary continuum extrapolation of the curvature yields 
$\kappa = 0.0132(18)$ for the chiral condensate and 
$\kappa = 0.0126(22)$ for the chiral susceptibility.}

This is in marginal agreement 
with recent results obtained by the method of
analytic continuation~\cite{ccp}, and larger 
than previous lattice determinations obtained by the Taylor
expansion technique \cite{Kaczmarek2011, Endrodi2011}
({notice however that larger values of the curvature
have been obtained when considering 
different observables, like the strange quark number susceptibility
or thermodynamical quantities}~\cite{Endrodi2011,Borsanyi2012}). {To better assess this issue}, {we have analyzed 
various possible systematic effects. Adopting the same
conventions for renormalizing the chiral condensate and for locating $T_c$ used in 
Ref.~\cite{Endrodi2011}, our estimate from the condensate 
would go down to $\kappa = 0.0110(18)$; hence, 
taking into account such systematics, 
the discrepancy with respect to results
with the same lattice discretization appears to be below the 
$2\, \sigma$ level.}
Adopting the disconnected chiral susceptibility
in place of the full renormalized one, as in Ref.~\cite{ccp}, 
our estimate from this observable would go 
up to $\kappa = 0.0146(41)$, in better agreement with Ref.~\cite{ccp}. 

We have also considered the effects of the introduction of a nonzero strange
quark chemical potential {{$\mu_s = \mu_l$.
The curvature stays unchanged within the present accuracy. This is 
reassuring for what concerns the comparison with the 
phenomenological conditions reproduced in 
heavy ion collisions: the requirement for strangeness neutrality
corresponds, around the transition, to $\mu_S/\mu_B \sim 1/4$
(see, e.g., Ref.~\cite{strangeness}), hence to
$\mu_s \sim \mu_l/4$ (see Eq.~(\ref{defchem1})), i.e. a value
of $\mu_s$ which is in-between the two cases explored in our simulations.}
On the contrary, non-linear terms in $\mu_l^2$  become larger
for $\mu_s = \mu_l$.  The
origin of this difference could be {related to} the different phase structure which is
found, for imaginary chemical potentials, {at} the different values of $\mu_s$.
In particular, the so-called Roberge-Weiss line and the associated non-analytic
behavior moves further from the $\mu_l = 0$ axis when $\mu_s = 0$.

{Given the agreement between the two determinations,
obtained from the renormalized chiral condensate and from the renormalized chiral susceptibility,
we quote $\kappa = 0.013(2)(1)$ as our present estimate
for the curvature of the chiral pseudo-critical line.
The second error is an estimate of the residual systematic uncertainty on the 
cut-off effects, and is obtained by comparing with an extrapolation which 
assumes a flat behavior with $N_t$, which is also compatible with our present 
results. 
In the future we plan to extend our investigation to lattices with 
at least $N_t = 10$,
in order to fully remove the systematics associated with the continuum
extrapolation.}

\acknowledgments

We thank F.~Becattini, L.~Cosmai, G.~Endrodi, P.~de Forcrand, F.~Karsch, M.~P.~Lombardo, A.~Papa and
A.~Patella for useful discussions.  
{FS received funding from the
European Research Council under the European Community’s Seventh
Framework Programme (FP7/2007-2013) ERC grant agreement No 279757}.
 FN acknowledges financial support from the EU under project Hadron
Physics 3 (Grant Agreement n. 283286).  This work was partially supported by
the INFN SUMA project.  Simulations have been performed on the BlueGene/Q Fermi
at CINECA (Project Iscra-B/EPDISIM), 
and on the CSN4 Zefiro cluster of the Scientific Computing
Center at INFN-PISA.

\appendix

\section{{Parameter sets and data at $T=0$}}
\label{app:t0}
The determination of the renormalized observables used to locate $T_c$ requires
the computation of the chiral condensate and of the chiral susceptility in the
zero temperature limit and at the same UV cutoff (\ie the same values of
$\beta$ and of the bare quark masses) of the finite temperature data. To that
aim, we spanned a range of $\beta$ on the line of constant physics, $3.5 \leq
\beta \leq 3.8$, on a symmetric $32^4$ lattice, corresponding to $T$ in the
range $25 - 55$ MeV.  We report the results for the condensate and for the
susceptibility in Table~\ref{tab:simT0}.

The temperatures are low enough to be considered as a good approximation 
of the $T = 0$ limit; indeed, as expected because of the absence of transitions in this $T$ 
range, one observes a very smooth dependence of the observables on $\beta$.
Hence, the relatively coarse sampling of the interval is enough
to allow {for} a reliable interpolation.
We adopted a cubic spline interpolation for the condensate and
a linar fit for the susceptibility.

The renormalization prescription for the susceptibility {that} we adopted throughout
the paper requires the subtraction of the $T=0$ result from the $T\neq0$ contribution.
To give an idea of the relative magnitude of the two contributions,
in Fig~\ref{susc_T0} we plot the chiral susceptibilities $\chi_{\bar{\psi}\psi}$,
defined in Eq.~(\ref{susc}), both at zero and at finite temperature, at zero chemical 
potential.

\begin{figure}[h!]
\includegraphics[width=0.92\columnwidth, clip]{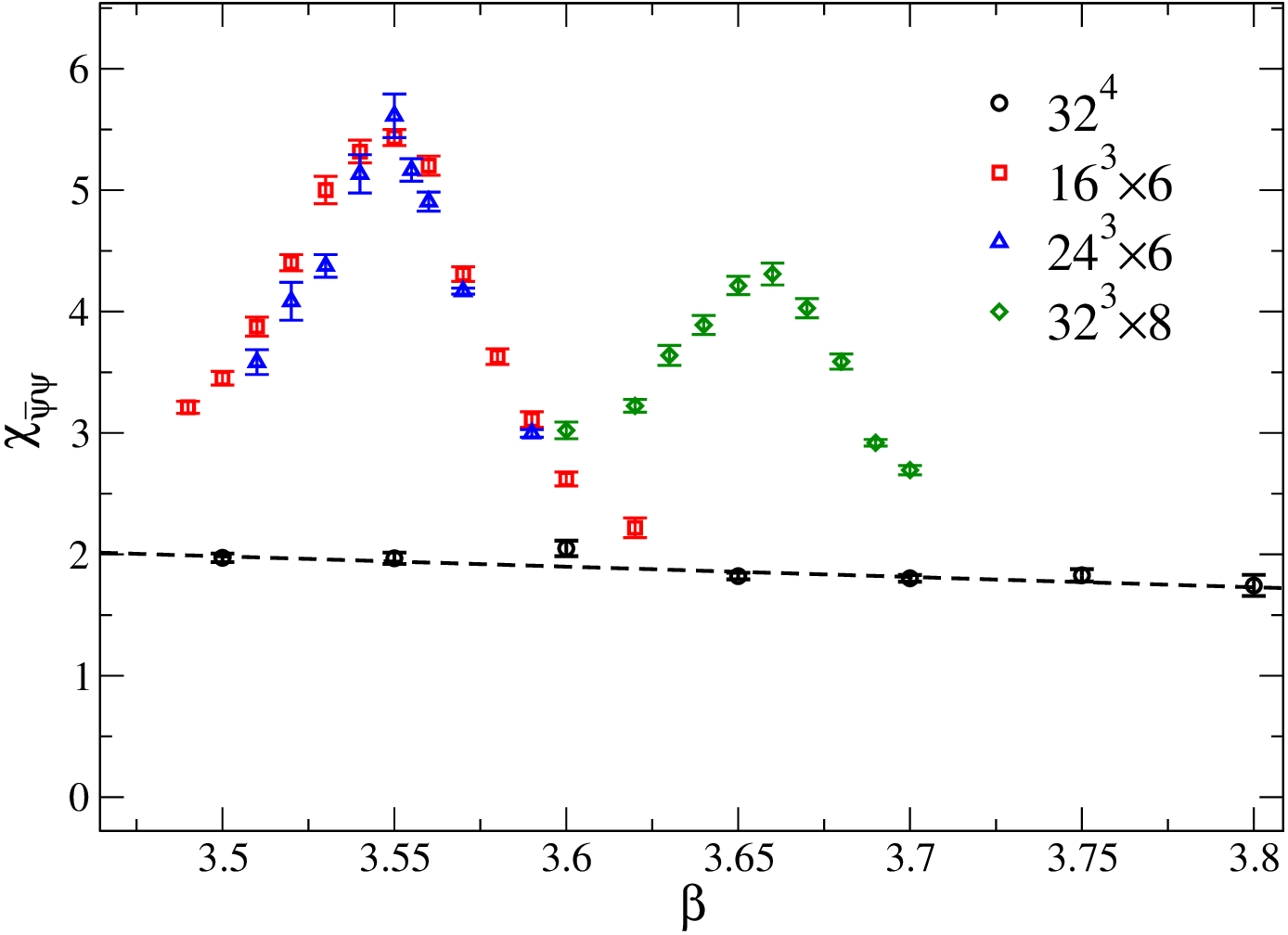}
\caption{Comparison between the zero and finite temperature chiral susceptibility {at zero chemical 
potential}. These terms are required to compute the renormalized chiral susceptibility, see Eq.~(\ref{rensusc}).
Data on the vertical axis are in lattice units. A linear fit of the $T = 0$ data
is also shown.}
\label{susc_T0}
\end{figure}

\begin{table}[t]
\centering
\begin{tabular}{ |c|c|c| }
\hline
$\beta$ & $\chi_{\bar{\psi}\psi}$ & \rule{0mm}{3.5mm}$\langle\bar{\psi}\psi\rangle - \ 2 (m_l / m_s)\langle\bar{s}s\rangle$ \\
\hline
3.50 & 1.972(35) & 0.07999(11) \\
3.55 & 1.968(46) & 0.05680(13) \\
3.60 & 2.049(64) & 0.03912(14) \\
3.65 & 1.821(27) & 0.02633(17) \\
3.70 & 1.803(28) & 0.018041(3) \\
3.75 & 1.827(51) & 0.01208(11) \\
3.80 & 1.744(88) & 0.008348(7) \\
\hline
\end{tabular}
\caption{Determination of the observables at $T=0$ (on the $32^4$ lattice)
needed to perform the renormalizations discussed in Section~\ref{sec2}. Data
are in lattice units.}
\label{tab:simT0}
\end{table}

\begin{table}[h]
\centering
\begin{tabular}{ |c|c|c|c|c|c| }
\hline
\rule{0mm}{3.5mm}$\beta$ & $L_s$ &  $N_t$ & $m_s$ & $a[\textnormal{fm}]$\\
\hline
3.490 & 16       & 6 & 0.132 &  0.2556  \\
3.500 & 16,32    & 6,32 & 0.126 &  0.2490  \\
3.510 & 16,24,32 & 6 & 0.121 &  0.2425  \\
3.520 & 16,24,32 & 6 & 0.116 &  0.2361  \\
3.530 & 16,24,32 & 6 & 0.111 &  0.2297  \\
3.540 & 16,24,32 & 6 & 0.10643 &  0.2235  \\
3.550 & 16,24,32 & 6,32 & 0.10200 &  0.2173  \\
3.555 & 24       & 6 & 0.09987 &  0.2142  \\
3.560 & 16,24,32 & 6 & 0.09779 &  0.2112  \\
3.570 & 16,24,32 & 6 & 0.09378 &  0.2052  \\
3.580 & 16,24,32 & 6 & 0.08998 &  0.1994  \\
3.590 & 16,24,32 & 6 & 0.08638 &  0.1937  \\
3.595 & 24       & 6 & 0.08465 &  0.1908  \\
3.600 & 16,24,32 & 6,8,32 & 0.08296 &  0.1881  \\
3.610 & 16,32    & 6 & 0.07973 &  0.1826  \\
3.620 & 16,24,32 & 6,8 & 0.07668 &  0.1773  \\
3.630 & 16,32    & 6,8 & 0.07381 &  0.1722  \\
3.640 & 16,32    & 6,8 & 0.07110 &  0.1672  \\
3.650 & 16,32    & 6,8,32 & 0.06854 &  0.1625  \\
3.660 & 16,32    & 6,8 & 0.06615 &  0.1579  \\
3.670 & 32       & 8 & 0.06390 &  0.1535  \\
3.680 & 16,32    & 6,8 & 0.06179 &  0.1493  \\
3.690 & 32       & 8 & 0.05982 &  0.1453  \\
3.700 & 32       & 8,32 & 0.05797 &  0.1416  \\
3.710 & 32       & 8 & 0.05624 &  0.1379  \\
3.720 & 32       & 8 & 0.05462 &  0.1345  \\
3.730 & 32       & 8 & 0.05309 &  0.1312  \\
3.740 & 32       & 8 & 0.05166 &  0.1280  \\
3.750 & 32       & 8,32 & 0.05030 &  0.1249  \\
3.800 & 32       & 32 & 0.04445 &  0.1110  \\
\hline
\end{tabular}
\caption{List of bare parameters and lattice spacings 
adopted in the simulations at zero and non-zero $T$;
the light quark mass is not reported, we have set
$m_l = m_s/28.15$ in all cases.
For each set of bare parameters, we report the various spatial
and temporal sizes adopted (notice that some possible combinations
of $L_s$ and $N_t$ have not been used). The 
systematic error on the lattice spacing is of the order of
2-3\,\%~\cite{tcwup1,befjkkrs}.}
\label{tab:parameters}
\end{table}

{Finally, for completeness, we report in Table~\ref{tab:parameters}
the full set of bare parameters and lattice spacings
adopted in our simulations, both at zero and non-zero $T$.
Data have been determined by a spline
interpolation of the values reported in Refs.}~\cite{tcwup1,befjkkrs}.
{As a crosscheck, we have performed an independent determination of the 
pion mass on our zero temperature lattices, obtaining values
in the range 133-137 MeV for the set of parameters explored: given 
the overall 2-3\% systematic error on the determination of the 
lattice spacing, this is satisfactory.}

\section{Comparison with other methods}
\label{appendix:comparisons}

\begin{table}[h]
\centering
\begin{tabular}{ |c|c|c|c|c|c|c| }
\hline
Lattice & $\theta_l/\pi$ & $\theta_s/\pi$ & $C_1M_1$ & $C_1M_2$ & $C_2M_1$ & $C_2M_2$\\
\hline
%\rule{0mm}{3mm}
$16^3\times 6$ & 0.00  & 0.00   & 148.2(3)   & 148.2(2)  &  148.4(4)  &  148.4(2) \\
$16^3\times 6$ & 0.20  & 0.00   & 155.0(4)   & 154.6(2)  &  155.1(5)  &  154.8(2) \\
$16^3\times 6$ & 0.24  & 0.00   & 158.9(4)   & 157.8(2)  &  159.1(4)  &  158.1(2) \\
$16^3\times 6$ & 0.275 & 0.00   & 161.2(4)   & 160.5(2)  &  161.5(4)  &  160.8(2) \\
\hline                                                                   
%\rule{0mm}{3mm}                                                                   
$24^3\times 6$ & 0.00 & 0.00    & 149.0(6)   & 149.0(2)  &  149.0(6)  &  149.0(2) \\
$24^3\times 6$ & 0.24 & 0.00    & 160.8(7)   & 159.6(2)  &  160.7(5)  &  159.6(2) \\
$24^3\times 6$ & 0.275 & 0.00   & 164.1(4)   & 163.0(2)  &  164.3(3)  &  163.1(2) \\
\hline                                                                   
%\rule{0mm}{3mm}                                                                   
$32^3\times 6$ & 0.00 & 0.00    & 149.3(3)   & 149.3(1)  & 149.4(4) &  149.4(1)  \\
$32^3\times 6$ & 0.24 & 0.00    & 160.2(2)   & 159.5(1)  & 160.4(2) &  159.6(1)  \\
$32^3\times 6$ & 0.275 & 0.00   & 163.5(3)   & 162.7(1)  & 163.5(3) &  162.7(1)  \\
\hline                                                                   
%\rule{0mm}{3mm}                                                                   
$32^3\times 8$ & 0.00 & 0.00    & 154.2(4)   & 154.2(2)  &  154.5(4)  &  154.5(2) \\
$32^3\times 8$ & 0.20 & 0.00    & 162.9(8)   & 161.6(2)  &  163.0(6)  &  161.8(2) \\
$32^3\times 8$ & 0.24 & 0.00    & 165.0(5)   & 164.5(2)  &  164.8(5)  &  164.5(2) \\
$32^3\times 8$ & 0.275 & 0.00   & 169.5(9)   & 168.6(3)  &  168.6(7)  &  168.4(3) \\
$32^3\times 8$ & 0.20 & 0.20    & 163.9(6)   & 163.3(2)  &  163.7(6)  &  163.4(2) \\
$32^3\times 8$ & 0.24 & 0.24    & 169.4(7)   & 168.3(3)  &  168.6(6)  &  168.3(3) \\
$32^3\times 8$ & 0.275 & 0.275  & 175.4(6)   & 174.1(2)  &  174.4(7)  &  174.0(3) \\
\hline
\end{tabular}
\caption{{Critical temperatures obtained by using different renormalization prescription and/or different
definition of $T_c$, see text for symbol definitions.}}
\label{tab:confronti_tc}
\end{table}

\begin{table}[h]
\centering
\begin{tabular}{ |c|c|c|c|c|c|c| }
\hline
Lattice & $\mu_s$ & Fit & $T_c(0)$ & $\kappa$ & $b$ & $\chi^2/n_{dof}$ \\
\hline
\rule{0mm}{3mm}
$16^3\times6$ & $ 0.00 $      & lin    &       $148.2(2)$    &         $0.0136(3)    $   & -                  & 0.8  \\
$24^3\times6$ & $ 0.00 $      & lin    &       $149.0(2)$    &         $0.0139(3)    $   & -                  & 0.2  \\
$32^3\times6$ & $ 0.00 $      & lin    &       $149.3(1)$    &         $0.0133(2)    $   & -                  & 0.1  \\
$32^3\times8$ & $ 0.00 $      & lin    &       $154.2(2)$    &         $0.0136(3)    $   & -                  & 2.5  \\
$32^3\times8$ & $ \mu_{l} $   & lin    &       $154.0(2)$    &         $0.0187(3)    $   & -                  & 15.5 \\
$32^3\times8$ & $ \mu_{l} $   & quad   &       $154.3(2)$    &         $0.0137(9)    $   & 0.0008(2)          & 0.02 \\
\hline
\end{tabular}
\caption{{Curvatures obtained by fitting the $T_c$s from the $C_1M_2$ combination of Table~\ref{tab:confronti_tc}.}}
\label{tab:C1M2}
\end{table}

\begin{table}[h]
\centering
\begin{tabular}{ |c|c|c|c|c|c|c| }
\hline
Lattice & $\mu_s$ & Fit & $T_c(0)$ & $\kappa$ & $b$ & $\chi^2/n_{dof}$ \\
\hline
\rule{0mm}{3mm}$16^3\times6$ & $ 0.00 $      & lin    &       $148.5(3)$    &         $0.0133(5)    $   & -          & 1.1 \\
$24^3\times6$ & $ 0.00 $      & lin    &       $149.1(5)$    &         $0.0152(7)    $   & -          & 0.0 \\
$32^3\times6$ & $ 0.00 $      & lin    &      $149.5(3)$    &         $0.0141(5)    $   & -          & 0.4 \\
$32^3\times8$ & $ 0.00 $      & lin    &       $154.7(4)$    &         $0.0135(7)    $   & -          & 2.5 \\
$32^3\times8$ & $ \mu_{l} $   & lin    &       $154.3(3)$    &         $0.0186(5)    $   & -          & 4.5 \\
$32^3\times8$ & $ \mu_{l} $   & quad   &       $154.3(3)$    &         $0.0138(11)   $   & 0.0008(3)  & 0.0 \\
\hline
\end{tabular}
\caption{{Curvatures obtained by fitting the $T_c$s from the $C_2M_1$ combination of Table~\ref{tab:confronti_tc}.}}
\label{tab:C2M1}
\end{table}

\begin{table}[h]
\centering
\begin{tabular}{ |c|c|c|c|c|c|c| }
\hline
Lattice & $\mu_s$ & Fit & $T_c(0)$ & $\kappa$ & $b$ & $\chi^2/n_{dof}$ \\
\hline
\rule{0mm}{3mm}$16^3\times6$ & $ 0.00 $        & lin   &       $148.33(16)$  &         $0.0124(3)  $   & -       & 12.5\\
$24^3\times6$ & $ 0.00 $        & lin   &       $148.49(23)$  &         $0.0147(3)  $   & -       & 0.0	\\
$32^3\times6$ & $ 0.00 $        & lin   &       $149.4(1)$  &         $0.0133(2)  $   & -       & 0.0	\\
$32^3\times8$ & $ 0.00 $        & lin   &       $154.55(17)$  &         $0.0131(3)  $   & -       & 1.7	\\
$32^3\times8$ & $ \mu_{l} $   & lin   &       $154.23(18)$  &         $0.0181(3)  $   & -       & 14.6\\
$32^3\times6$ & $ \mu_{l} $   & quad  &       $154.56(19)$  &         $0.0133(10) $   &0.0006(8)& 0.0 \\
\hline
\end{tabular}
\caption{{Curvatures obtained by fitting the $T_c$s from the $C_2M_2$
combination (\ie the same method adopted in Ref.~\cite{Endrodi2011}) of
Table~\ref{tab:confronti_tc}.}}
\label{tab:C2M2}
\end{table}

\begin{table}[t!]
\centering
\begin{tabular}{ |c|c|c|c|c| }
\hline
Lattice & $\theta_l/\pi$ & $\theta_s/\pi$ & $T_c(\chi^r)$ & $T_c(\chi_{disc}) $ \\
\hline
\rule{0mm}{3mm}$16^3\times 6$ & 0.00 & 0.00    & 150.7(4)   &  145.8(7)  \\ 
$16^3\times 6$ & 0.20 & 0.00    & 157.0(4)   &  151.9(9)  \\
$16^3\times 6$ & 0.24 & 0.00    & 160.0(4)   &  155.6(9)  \\
$16^3\times 6$ & 0.275 & 0.00   & 162.7(4)   &  158.0(7)  \\
\hline
\rule{0mm}{3mm}$24^3\times 6$ & 0.00 & 0.00    & 151.6(5)   &  148.0(1.0)\\
$24^3\times 6$ & 0.24 & 0.00    & 162.0(5)   &  158.3(8)  \\
$24^3\times 6$ & 0.275 & 0.00   & 165.9(4)   &  162.2(9)  \\
\hline
\rule{0mm}{3mm}$32^3\times 6$ & 0.00 & 0.00    & 152.0(4)   &  147.2(1.0)\\
$32^3\times 6$ & 0.24 & 0.00    & 162.7(4)   &  156.9(9)  \\
$32^3\times 6$ & 0.275 & 0.00   & 165.5(4)   &  161.5(1.3)  \\
\hline
\rule{0mm}{3mm}$32^3\times 8$ & 0.00 & 0.00    & 155.6(7) &  151.2(1.2)\\
$32^3\times 8$ & 0.20 & 0.00    & 163.0(6) &  157.2(1.0)\\
$32^3\times 8$ & 0.24 & 0.00    & 165.8(8) &  160.4(1.4)\\
$32^3\times 8$ & 0.275 & 0.00   & 169.8(7) &  166.1(1.3)\\
$32^3\times 8$ & 0.20 & 0.20    & 165.3(9) &  159.3(9)  \\
$32^3\times 8$ & 0.24 & 0.24    & 169.6(7)   &  164.8(1.5)\\
$32^3\times 8$ & 0.275 & 0.275  & 177.0(8)   &  172.9(1.3)\\
\hline
\end{tabular}
\caption{Critical temperatures obtained from the non-renormalized disconnected chiral susceptibility. 
The values obtained from $\chi^r$ are reported for reference.}
\label{tab:tc_susc_disconnected}
\end{table}

\begin{table}[h!]
\centering
\begin{tabular}{ |c|c|c|c|c|c|c| }
\hline
Lattice & $\mu_s$ & Fit & $T_c(0)$ & $\kappa$ & $b$ & $\chi^2/n_{dof}$ \\
\hline
\rule{0mm}{3mm}$16^3\times6$ & $ 0.00 $      & lin   &   $145.8(7)$  &  $0.0126(10)  $   & -       & 0.2\\
$24^3\times6$ & $ 0.00 $      & lin   &   $147.9(1.0)$  &  $0.0141(13)  $   & -       & 0.2\\
$32^3\times6$ & $ 0.00 $      & lin   &   $147.0(1.0)$  &  $0.0138(16)  $   & -       & 0.8\\
$32^3\times8$ & $ 0.00 $      & lin   &   $150.5(1.1)$  &  $0.0143(17)  $   & -       & 1.3\\
$32^3\times8$ & $ \mu_{l} $   & lin   &   $149.8(1.1)$  &  $0.0208(18)  $   & -       & 3.8\\
$32^3\times8$ & $ \mu_{l} $   & quad  &   $151.2(1.2)$  &  $0.008(5)    $   &0.0020(7)& 0.04\\
\hline
\end{tabular}
\caption{{Curvatures extracted from the data of Table~\ref{tab:tc_susc_disconnected}.}}
\label{tab:fit_susc_disconnected}
\end{table}

In this Appendix we analyze how the determinations of $T_c$
and $\kappa$ change if different prescriptions are adopted
to renormalize observables or to locate $T_c$.

Regarding the determination of $\kappa$ from the chiral condensate
we compare two different renormalization prescriptions of the
condensate and two different methods to extract $T_c$.
We set the following notation: $C_1$ is the chiral condensate
renormalized as in Eq.~(\ref{rencond}), while $C_2$
is the one renormalized as in Eq.~(\ref{eq:ren_pres_wupp}).
For {what concerns} the method we define $M_1$ as {the determination of} $T_c$ {obtained from} the inflection point of $\langle\bar{\psi}\psi\rangle^r_l$,
which is the one adopted in this work, and $M_2$ as the
{determination} based on Eq.~(\ref{eq:tc_mu_wupp}); in the latter case,
we have taken the value of the condensate at the inflection point 
at zero chemical potential as the reference value which is kept
constant at nonzero $\mu_B$.

According to these definitions, the method that we adopted throughout
the paper can be addressed as $C_1M_1$.
In Table~\ref{tab:confronti_tc} we show the results for $T_c$
obtained by taking all different combinations.
As we did in Section~\ref{sec4}, we compute the curvature of the pseudo-critical line
by fitting the extracted values of $T_c$ by Eq.~(\ref{fitfun}).
The results of these fits are reported in Table~\ref{tab:C1M2},~\ref{tab:C2M1}
and ~\ref{tab:C2M2} respectively for the combinations $C_1M_2$, $C_2M_1$ and $C_2M_2$.

Regarding the determination of $\kappa$ from the chiral susceptibility,
in Table~\ref{tab:tc_susc_disconnected} we report the results for the pseudo-critical
temperature as a function of the chemical potential obtained from
the non-renormalized disconnected chiral susceptibility $\chi^{disc}_{\bar{\psi}\psi}$,
defined in Eq.~(\ref{sconn}). {The disconnected susceptibility 
is measured in lattice spacing units: that leads to an additional 
$T = 1/(N_t a)$ dependence, which can account for the generally lower
values of $T_c$ obtained.}
Again, as we did in Section~\ref{sec4}, we compute the curvature of the pseudo-critical line
by fitting $T_c$ with Eq.~(\ref{fitfun}): we report the results in Table~\ref{tab:fit_susc_disconnected}.

\section{One loop Polyakov effective potential}
\label{appendix:effective_poly}

The effective potential for the Polyakov loop can be computed in perturbation
theory, obtaining results {valid} in the {limit of very} high 
{temperature.}
The effective potential
can be written in term of the eigenvalues of the Polyakov loop, denoted by
$\lambda_j=e^{i\phi_j}$ (with $\sum_j \phi^j= 0\ \mathrm{mod\ } 2\pi$). A one-loop 
computation in an $SU(N_c)$ gauge theory coupled to $N_f$ massless
fermions gives (see, e.g., Ref.~\cite{RW})
\begin{equation}\label{poly_pot}
V_{\mathrm{eff}}=\frac{\pi^2 T^4}{12}\left(\frac{1}{2} \sum_{j,k=1}^{N_c} V^{(g)}(j,k)
-\sum_{f=1}^{N_f}\sum_{j=1}^{N_c} V^{(q)}(f, j)\right)
\end{equation}
where $V^{(g)}$ and $V^{(q)}$ are the gluon and quark contributions
respectively, explicitly given by 
\begin{equation}
\begin{aligned}
& V^{(g)}(j,k) = \left(1-\left(\left[\frac{\phi_{j}}{\pi}-\frac{\phi_{k}}{\pi}\right]_{\mathrm{mod\ 2}}
-1\right)^{2}\right)^{2} \\
& V^{(q)}(f, j) = \left(1-\left(\left[\frac{\phi_{j}+ \theta_f}{\pi}+1\right]_{\mathrm{mod\ 2}}-1\right)^{2}
\right)^{2} . 
\end{aligned}
\end{equation}
In the expression for $V^{(q)}$, $\theta_f=\mathrm{Im}(\mu_f)/T$ is the angle
introduced in Section~\ref{sec2}, associated with the chemical potential of the
quark flavor $f$. 

{For $SU(3)$ pure gauge theory the minima of $V_{\mathrm{eff}}$ are degenerate and
located on the cubic roots of the identity, i.e. $\phi_j=\frac{2\pi}{3}k\ \forall j$, where
$k=0,1,2$. Fermions lift this degeneracy: the cubic roots are still 
local minima but they are not degenerate any more; in the case of
three flavors having the same
chemical potential (i.e. $\mu_u=\mu_d=\mu_s$), the absolute
minimum changes by increasing $\theta$, with
transitions taking place at $\theta_c=(2 n+1)\pi/3$, where $n$ is an integer.
}

As noted in Section~\ref{sec3}, 
one expects the position of these transitions to
change when $\mu_u=\mu_d$ but $\mu_s=0$. This can be explicitly seen by using
Eq.~\eqref{poly_pot}. The potential evaluated on the three cubic roots of the
identity is plotted in Fig.~\ref{fig:potential}: the level
crossings (corresponding to sector changes) {move} with respect to the
$\mu_l=\mu_s$ setup, the first one being located at $\theta_c\approx 0.482933\, \pi$. 
{For comparison, we report also the analysis for the standard
RW case ($\mu_s = \mu_u = \mu_d$) in Fig.~\ref{fig:potential_rw}.}
\\
 
\begin{figure}[h!]
\includegraphics[width=0.9\columnwidth, clip]{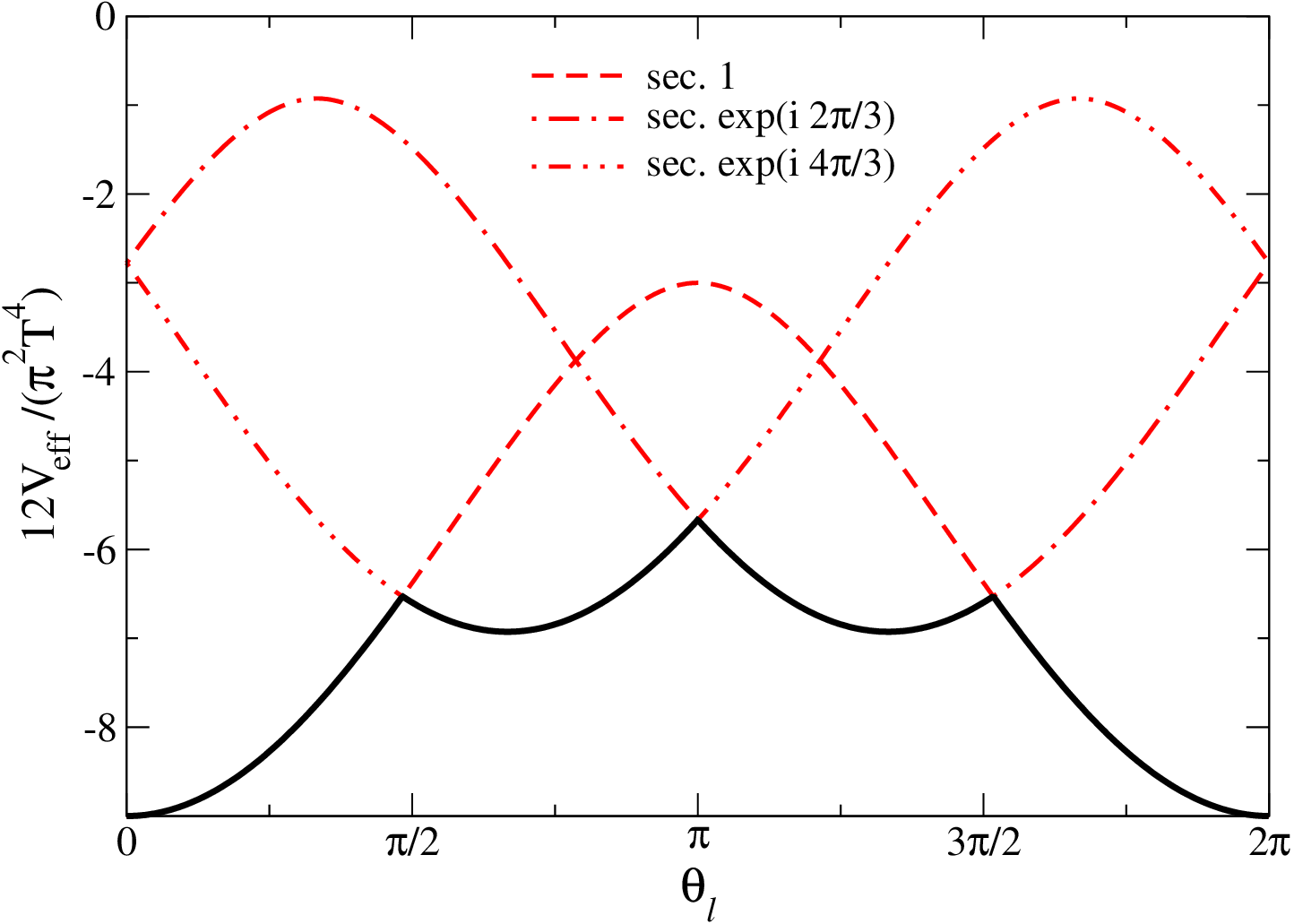}
\caption{Effective Polyakov loop potential computed from Eq.~\eqref{poly_pot} for $\mu_u=\mu_d$ and $\mu_s=0$.} 
\label{fig:potential}
\end{figure}

\begin{figure}[h!]
\includegraphics[width=0.9\columnwidth, clip]{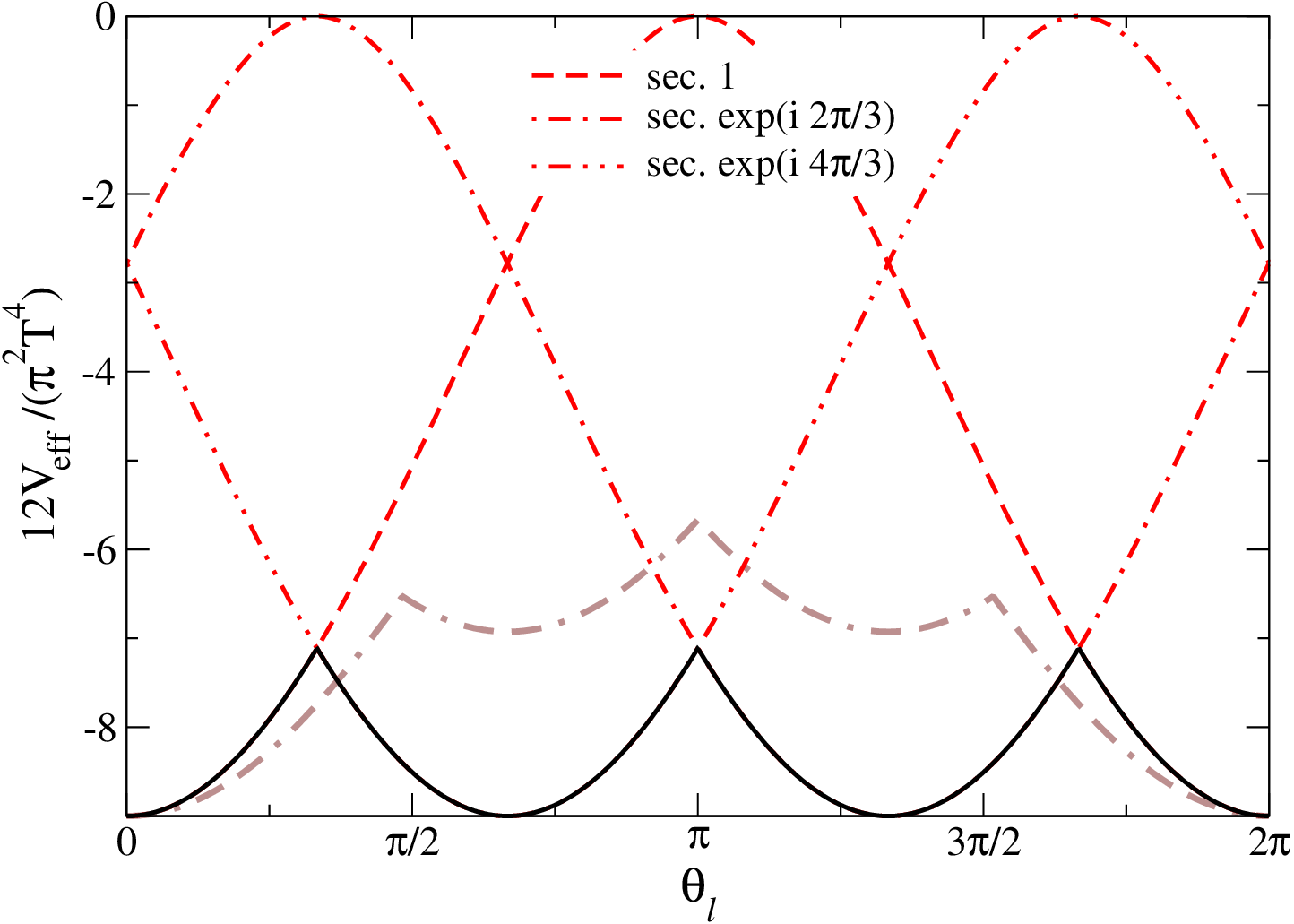}
\caption{Same as in Fig.~\ref{fig:potential}, but for 
$\mu_s = \mu_u = \mu_d$:
this is actually the same as Fig.~2 of Ref.~\cite{RW}.
The true vacuum potential from Fig.~\ref{fig:potential} is also 
reported to allow for a direct comparison of the two cases.}
\label{fig:potential_rw}
\end{figure}

\end{document}